\documentclass[oneside,a4paper,11pt,explicit]{book}
\def\ARXIVBUILD{1}
\ifdefined\ARXIVBUILD
\else
  \csname input\endcsname{arxiv_volume_V.tex}
\fi

\usepackage[T1]{fontenc}
\usepackage[utf8]{inputenc}
\usepackage{textcomp}
\usepackage{vol5_kultem}

\usepackage{amsmath,amssymb}
\usepackage{booktabs}
\usepackage{tabularx}
\usepackage{array}
\usepackage{longtable}
\usepackage{lscape}
\usepackage{graphicx}
\usepackage[dvipsnames,svgnames,table]{xcolor}
\usepackage[most]{tcolorbox}
\usepackage{float}
\usepackage{hyperref}

\makeatletter
\renewcommand*\l@section{\@dottedtocline{1}{1.5em}{2.8em}}
\renewcommand*\l@subsection{\@dottedtocline{2}{3.8em}{3.7em}}
\renewcommand*\l@subsubsection{\@dottedtocline{3}{7.0em}{4.6em}}
\makeatother

\usetikzlibrary{arrows.meta,calc,positioning,shapes.geometric,
  decorations.pathmorphing,backgrounds,fit}
\numberwithin{equation}{chapter}

\definecolor{ink}{HTML}{151B23}
\definecolor{titlebg}{HTML}{100880}
\definecolor{softink}{HTML}{263441}
\definecolor{muted}{HTML}{5C6875}
\definecolor{paper}{HTML}{F6F7F2}
\definecolor{panel}{HTML}{FFFFFF}
\definecolor{line}{HTML}{D8DEE5}
\definecolor{teal}{HTML}{007C77}
\definecolor{blue}{HTML}{245AA6}
\definecolor{gold}{HTML}{B86B00}
\definecolor{rose}{HTML}{A7354D}
\definecolor{green}{HTML}{23724A}
\definecolor{violet}{HTML}{5A4CA0}

\newcommand{\facility}{3.5-meter Segmented-Mirror Robotic Space Telescope}

\newcommand{\kms}{km\,s$^{-1}$}

\newcolumntype{Y}{>{\raggedright\arraybackslash}X}

\newtcolorbox{leadbox}[2][]{
  enhanced, colback=paper, colframe=#2, boxrule=0.9pt, arc=2mm,
  left=2.2mm, right=2.2mm, top=1.8mm, bottom=1.8mm,
  fonttitle=\sffamily\bfseries, coltitle=white,
  attach boxed title to top left={xshift=2mm,yshift=-2mm},
  boxed title style={colback=#2,arc=1.2mm,boxrule=0pt}, #1 }
\newtcolorbox{metricbox}[1]{
  enhanced, colback=white, colframe=#1, boxrule=0.65pt, arc=1.3mm,
  left=1.8mm, right=1.8mm, top=1.6mm, bottom=1.6mm }

\makeatletter
\renewenvironment{thebibliography}[1]
  {\section*{References}\@mkboth{}{}%
   \list{\@biblabel{\@arabic\c@enumiv}}%
        {\settowidth\labelwidth{\@biblabel{#1}}%
         \leftmargin\labelwidth \advance\leftmargin\labelsep
         \usecounter{enumiv}\let\p@enumiv\@empty
         \renewcommand\theenumiv{\@arabic\c@enumiv}}%
   \small\sloppy\clubpenalty4000\widowpenalty4000\sfcode`\.\@m}
  {\def\@noitemerr{\@latex@warning{Empty `thebibliography' environment}}\endlist}
\makeatother

\newcommand{\wpvolumelabel}{V}
\newcommand{\wpvolumetitle}{Key Scientific Mission: Compact-Object Time-Domain Science}
\title{3.5-meter Segmented-Mirror Robotic Space Telescope: \wpvolumelabel. \wpvolumetitle}
\subtitle{Mission White Paper: \wpvolumelabel. \wpvolumetitle}
\newcommand{\wpauthors}{Juhan Kim$^{1}$, Yong-Woo Kang$^{2}$, Sang Hyun Lee$^{2,3}$, Jeong-Yeol Han$^{2,4}$, Sungwook E. Hong$^{2,4}$, Bongkon Moon$^{2,4}$, Donguk Song$^{2}$, Juhyung Kang$^{2}$, Myeong-Gu Park$^{5}$, Sang Chul Kim$^{2,4}$, Chung-Uk Lee$^{2}$, Sangmo Tony Sohn$^{6}$, Arman Shafieloo$^{2,4}$, David Parkinson$^{2,4}$, Hong Soo Park$^{2,4}$, Dohyeong Kim$^{7}$, Chan Park$^{2}$, Jungjoo Sohn$^{8}$, Young-Beom Jeon$^{2}$, Jong-Hak Woo$^{9}$, Hyung Mok Lee$^{9}$, Hong Bae Ann$^{7}$, Myungkook James Jee$^{10}$, Mansoo Choi$^{2}$, Changbom Park$^{1}$}
\date{2026}
\newcommand{\wpabstract}{%
An isolated compact object retains the point-source resolving power of the space-based slitless spectrograph. The baseline wavelength range is $0.2$--$1.5\,\mu\mathrm{m}$. The planning baseline uses $R\simeq1000$ for broad and faint transient spectra and reserves selectable bands at $R\simeq5000$ for accretion-disk profiles, velocity structure, and precision line ratios. Broad features can be measured after binning the native $R\simeq5000$ data to lower resolution. Rapid-response spectroscopy follows gravitational-wave counterparts and kilonovae from hours to days. Repeated spectra of dwarf novae and compact binaries trace accretion state and orbital phase, while uninterrupted imaging of white dwarfs measures pulsation frequencies. The program combines mission-based monitoring with external alerts, including KGMT transient detections. The instrument study must preserve calibrated throughput to $2.70\,\mu\mathrm{m}$ and evaluate a $3.0\,\mu\mathrm{m}$ operational band edge, with $2.5\,\mu\mathrm{m}$ retained as the formal engineering off-ramp. Mid-infrared imaging is not part of the adopted compact-object baseline.
}
\newcommand{\wpkeywords}{\textbf{Keywords:} compact objects, time-domain astronomy, kilonovae, dwarf novae, white dwarfs, slitless spectroscopy}
\newcommand{\wpchapteroffset}{4}
\newcommand{\MakeVolumeBody}{%
  \definecolor{Blue1}{HTML}{D98A9A}%
  \definecolor{Blue2}{HTML}{B34F68}%
  \definecolor{Blue3}{HTML}{7A2944}%
}

\newcommand{\MakeFrontCover}{%
  \begin{titlepage}
  \thispagestyle{empty}\sffamily\centering
  \noindent\colorbox{Blue3}{\parbox[t]{\dimexpr\textwidth-2\fboxsep\relax}{%
    \vspace{4mm}\centering
    {\color{white}\bfseries\fontsize{24}{29}\selectfont 3.5-meter Segmented-Mirror\\[1.5mm]
      Robotic Space Telescope}\\[3.5mm]
    {\color{white}\Large Mission White Paper}\\[1.5mm]
    {\color{white!88}\large \wpvolumelabel. \wpvolumetitle}%
    \vspace{4mm}}}
  \vfill
  \includegraphics[width=\textwidth]{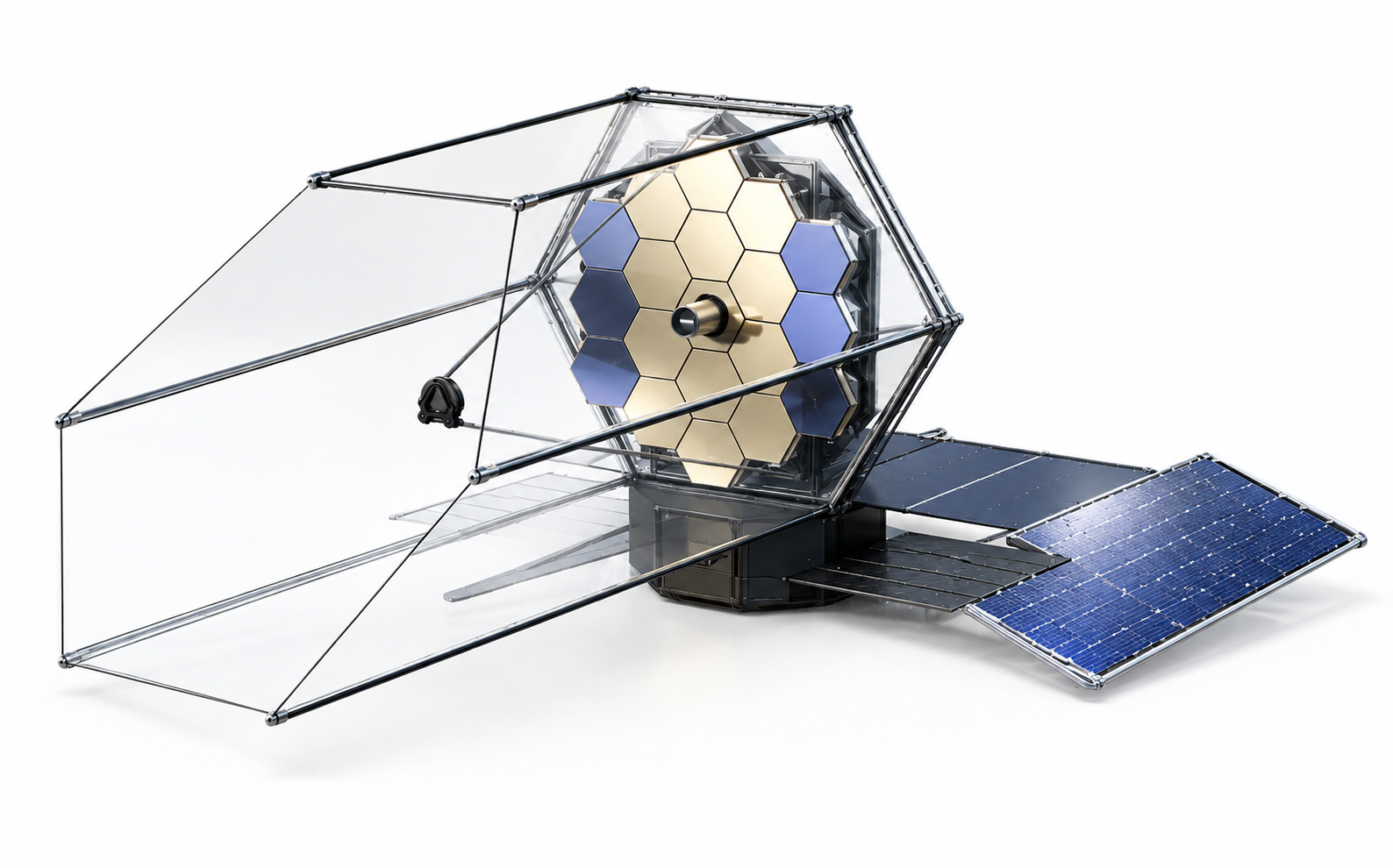}
  \vfill
  {\color{ink}\bfseries\normalsize \wpauthors\par}
  \vspace{3mm}
  {\color{muted}\small
    $^{1}$ Korea Institute for Advanced Study \textperiodcentered\
    $^{2}$ Korea Astronomy and Space Science Institute \\
    $^{3}$ University of Ulsan \textperiodcentered\
    $^{4}$ University of Science and Technology\\
    $^{5}$ Kyungpook National University \textperiodcentered\
    $^{6}$ Space Telescope Science Institute \textperiodcentered\
    $^{7}$ Pusan National University\\
    $^{8}$ Korea National University of Education \textperiodcentered\
    $^{9}$ Seoul National University \textperiodcentered\
    $^{10}$ Yonsei University\par}
  \vspace{5mm}
  {\color{Blue3}\bfseries\Large 2026}
  \vspace{4mm}
  \end{titlepage}}

\newcommand{\MakeColophon}{%
  \clearpage
  \thispagestyle{empty}\sffamily
  \null\vfill
  \noindent{\color{muted}\footnotesize
    {\color{ink}\bfseries 3.5-meter Segmented-Mirror Robotic Space Telescope}\\
    Mission White Paper: \wpvolumelabel. \wpvolumetitle\\[5mm]
    {\color{rose}\bfseries Release date September 17, 2026\\[5mm]}
    \textcopyright\ 2026 Korea Astronomy and Space Science Institute and the
    authors. All rights reserved.\\[2mm]
    Prepared by the mission study team at the Korea Astronomy and Space Science
    Institute, University of Ulsan, University of Science and Technology, the
    Space Telescope Science Institute, the Korea Institute for Advanced Study,
    Pusan National University, Korea National University of Education, Seoul
    National University, Kyungpook National University, and Yonsei University.\\[5mm]
    {\color{ink}\bfseries Suggested citation:} Kim, J., Kang, Y.-W., Lee, S.-H.,
    et al.\ (2026), \textit{3.5-meter Segmented-Mirror Robotic Space Telescope:
    Mission White Paper, \wpvolumelabel. \wpvolumetitle}.\\[2mm]
    {\color{ink}\bfseries Corresponding author:} Sang Hyun Lee
    \textperiodcentered\ \texttt{shlee@kasi.re.kr}.\par}
  \vspace{8mm}
  \clearpage}

\newcommand{\MakeBackCover}{%
  \clearpage
  \ifodd\value{page} \hbox{}\thispagestyle{empty}\newpage \fi
  \thispagestyle{empty}\null
  \clearpage
  \thispagestyle{empty}\sffamily\centering
  \null\vspace*{\stretch{1}}
  \includegraphics[width=\textwidth]{vol5_3.5ST_GPT.png}
  \vspace*{\stretch{2.2}}
  \noindent\colorbox{Blue3}{\parbox[t]{\dimexpr\textwidth-2\fboxsep\relax}{%
    \vspace{3mm}\centering
    {\color{white}\bfseries\Large 3.5-meter Segmented-Mirror Robotic Space Telescope}\\[1.5mm]
    {\color{white!88}\normalsize Mission White Paper: \wpvolumelabel. \wpvolumetitle}%
    \vspace{3mm}}}
  \vspace{3mm}
  \clearpage}

\hypersetup{
  colorlinks=true,
  linkcolor=blue,
  citecolor=teal,
  urlcolor=gold,
  pdftitle={3.5-meter Segmented-Mirror Robotic Space Telescope -- \wpvolumelabel. \wpvolumetitle},
  pdfauthor={Juhan Kim et al.}}

\begin{document}
\frontmatter
\MakeFrontCover
\MakeColophon
\thispagestyle{fancy}
\vspace*{0.5em}
\noindent{\Large\bfseries Abstract}\par\smallskip
\noindent\wpabstract\par\medskip
\noindent\wpkeywords
\clearpage
\tableofcontents

\mainmatter
\renewcommand{\chaptername}{Volume}
\renewcommand{\thechapter}{\Roman{chapter}}
\setcounter{chapter}{\wpchapteroffset}
\MakeVolumeBody
% Book bridge for Appendix X.  The source remains independently compilable.
\def\APPENDIXXBOOK{1}
%\graphicspath{{appendix_x_assets/}}
\providecommand{\Rpoint}{R_{\rm point}}
\providecommand{\micron}{\,\mu\mathrm{m}}
\renewcommand{\kms}{\mathrm{km\,s^{-1}}}
\newcolumntype{P}[1]{>{\raggedright\arraybackslash}p{#1}}
\newcolumntype{C}[1]{>{\centering\arraybackslash}m{#1}}
\providecommand{\topcell}[1]{%
  \begin{minipage}[t]{\linewidth}\vspace{0pt}#1\end{minipage}}
\newtcolorbox{xresultbox}[1]{
  enhanced,
  colback=white,
  colframe=#1,
  coltext=ink,
  boxrule=0.75pt,
  arc=1.5mm,
  left=2mm,right=2mm,top=1.6mm,bottom=1.6mm
}
\chapter{Compact-Object Time-Domain Science}
\label{app:compact-objects}
% !TEX program = xelatex
% Standalone Appendix D
% Compact-object time-domain science with selectable R=5000 slitless spectroscopy.

\ifdefined\APPENDIXXBOOK
% The book supplies the document class and shared packages.
\else
\documentclass[10pt,onecolumn]{article}

\usepackage[a4paper,margin=18mm,top=17mm,bottom=18mm]{geometry}
\usepackage{microtype}
\usepackage{amsmath,amssymb}
\usepackage{siunitx}
\usepackage{booktabs}
\usepackage{tabularx}
\usepackage{array}
\usepackage{longtable}
\usepackage{graphicx}
%\graphicspath{{vol5_}}
\usepackage[dvipsnames,svgnames,table]{xcolor}
\usepackage[most]{tcolorbox}
\usepackage{titlesec}
\usepackage{enumitem}
\usepackage{fancyhdr}
\usepackage{hyperref}
\usepackage{caption}
\usepackage{float}
\usepackage{tikz}
\usetikzlibrary{arrows.meta,calc,positioning,shapes.geometric,fit}

\definecolor{ink}{HTML}{151B23}
\definecolor{titlebg}{HTML}{5A284A}
\definecolor{softink}{HTML}{263441}
\definecolor{muted}{HTML}{5C6875}
\definecolor{paper}{HTML}{F7F6F2}
\definecolor{panel}{HTML}{FFFFFF}
\definecolor{line}{HTML}{D8DEE5}
\definecolor{teal}{HTML}{007C77}
\definecolor{blue}{HTML}{245AA6}
\definecolor{gold}{HTML}{B86B00}
\definecolor{rose}{HTML}{A7354D}
\definecolor{violet}{HTML}{5A4CA0}
\definecolor{orange}{HTML}{B55220}

\hypersetup{
  colorlinks=true,
  linkcolor=blue,
  citecolor=teal,
  urlcolor=gold,
  pdftitle={Volume V: Compact-Object Time-Domain Science},
  pdfauthor={Juhan Kim}
}

\pagestyle{fancy}
\fancyhf{}
\fancyhead[L]{\sffamily\footnotesize 3.5-meter Segmented-Mirror Robotic Space Telescope}
\fancyhead[R]{\sffamily\footnotesize Volume V: Compact Objects}
\fancyfoot[C]{\sffamily\footnotesize V-\thepage}
\renewcommand{\headrulewidth}{0.35pt}

\titleformat{\section}{\sffamily\Large\bfseries\color{ink}}{\thesection}{0.6em}{}
\titleformat{\subsection}{\sffamily\large\bfseries\color{violet}}{\thesubsection}{0.6em}{}
\titleformat{\subsubsection}{\sffamily\normalsize\bfseries\color{softink}}{\thesubsubsection}{0.6em}{}
\titlespacing*{\section}{0pt}{1.1em}{0.45em}
\titlespacing*{\subsection}{0pt}{0.85em}{0.3em}
\captionsetup{font={small,sf},labelfont={bf,color=violet}}
\setlist[itemize]{leftmargin=*,itemsep=2pt,topsep=3pt}
\setlist[enumerate]{leftmargin=*,itemsep=2pt,topsep=3pt}
\renewcommand{\arraystretch}{1.18}
\renewcommand{\thesection}{V.\arabic{section}}
\renewcommand{\thesubsection}{V.\arabic{section}.\arabic{subsection}}
\renewcommand{\thesubsubsection}{V.\arabic{section}.\arabic{subsection}.\arabic{subsubsection}}
\renewcommand{\thefigure}{V.\arabic{figure}}
\renewcommand{\thetable}{V.\arabic{table}}
\renewcommand{\theequation}{V.\arabic{equation}}
\newcolumntype{Y}{>{\raggedright\arraybackslash}X}
\newcolumntype{C}[1]{>{\centering\arraybackslash}m{#1}}
\newcolumntype{P}[1]{>{\raggedright\arraybackslash}p{#1}}
\newcommand{\topcell}[1]{\begin{minipage}[t]{\linewidth}\vspace{0pt}#1\end{minipage}}
\emergencystretch=3em

\newtcolorbox{leadbox}[1]{
  enhanced,
  colback=paper,
  colframe=#1,
  coltext=ink,
  boxrule=0.9pt,
  arc=2mm,
  left=2.2mm,right=2.2mm,top=1.8mm,bottom=1.8mm
}
\newtcolorbox{xresultbox}[1]{
  enhanced,
  colback=white,
  colframe=#1,
  coltext=ink,
  boxrule=0.75pt,
  arc=1.5mm,
  left=2mm,right=2mm,top=1.6mm,bottom=1.6mm
}

\newcommand{\facility}{3.5-meter Segmented-Mirror Robotic Space Telescope}
\newcommand{\Rpoint}{R_{\rm point}}
\newcommand{\kms}{\mathrm{km\,s^{-1}}}
\newcommand{\micron}{\,\mu\mathrm{m}}

\fi

\ifdefined\APPENDIXXBOOK
\let\appendixXfinish\relax
\else
\def\appendixXfinish{\end{document}}
\fi

\ifdefined\APPENDIXXBOOK
\else
\begin{document}
\thispagestyle{empty}

\begin{tcolorbox}[enhanced,colback=titlebg,colframe=titlebg,arc=2.5mm,
  left=5mm,right=5mm,top=7mm,bottom=7mm]
{\sffamily\color{white}\fontsize{23}{28}\selectfont\bfseries
Volume V: Compact-Object Time-Domain Science}\\[2.2mm]
{\sffamily\color{white}\fontsize{14}{18}\selectfont
with $R\simeq1000$ baseline and selectable $R\simeq5000$ spectroscopy}\\[3mm]
{\sffamily\color{white!80}\normalsize
Neutron-star and black-hole transients, white dwarfs, and dwarf novae}
\end{tcolorbox}
\fi

\vspace{2mm}
\noindent
\begin{tabularx}{\textwidth}{@{}P{38mm}Y@{}}
\topcell{\bfseries\color{violet}Baseline mode} &
\topcell{Point-source slitless spectroscopy with $R\simeq1000$ as the baseline
and selectable $R\simeq5000$ bands, supplemented by multi-band imaging and
high-cadence time series.}\\
\topcell{\bfseries\color{violet}Primary science} &
\topcell{Rapid optical and near-infrared follow-up of gravitational-wave
counterparts, plus accretion-disk spectroscopy of white dwarfs, neutron
stars, black holes, and dwarf novae.}\\
\topcell{\bfseries\color{violet}What is measured} &
\topcell{Line centroids, widths, asymmetries, equivalent widths, continuum
color, fading rate, orbital modulation, and spectral evolution.}\\
\topcell{\bfseries\color{violet}What is not claimed} &
\topcell{The telescope does not measure X-rays or gamma rays. A single
optical/NIR spectrum does not determine a neutron-star equation of state
without external gravitational-wave and radiative-transfer information.}
\end{tabularx}

\section{Science Scope and Instrument Logic}

Recent mission concepts have emphasized rapid access to transient and
multi-messenger sources as a major space-observatory capability
\cite{roy2026V,wevers2026V}. This volume defines a narrower and testable
compact-object program for the 3.5ST. It measures spectral evolution in
selected merger counterparts, velocity-resolved line profiles in accreting
binaries, and pulsation frequencies in white dwarfs. Each source class has a
specified cadence, signal-to-noise requirement, and analysis product. The
program therefore uses external alerts as one scheduling input while making
repeated calibrated optical and near-infrared spectroscopy its primary
measurement.

The scope is defined by a calibrated optical and near-infrared spectroscopic
program with alert-triggered observations as one scheduling input.
The adopted baseline is optical and near-infrared spectroscopy with a wide
survey setting near $R=1000$ and selectable bands near $R=5000$. The higher
resolution bands resolve double-peaked disk lines and line asymmetries in
dwarf novae and other accreting systems. They do not turn the observatory
into an X-ray or gamma-ray facility, and the program does not claim to infer
a neutron-star equation of state from one spectrum. Those inferences require
external gravitational-wave, X-ray, and radiative-transfer constraints.

Compact-object angular sizes are normally far below the diffraction scale.
An isolated compact object therefore retains the point-source resolution of
the slitless spectrograph. Extended-galaxy morphology instead broadens the
dispersed image. Point-like morphology does not remove spectral confusion.
Each target produces a long trace that can overlap the trace of an unrelated
field source. The measurement therefore requires direct imaging, a model of
every trace in the field, and more than one dispersion orientation.

The baseline is intentionally restricted to optical and near-infrared
observations. Kilonovae, accretion disks, dwarf novae, and polluted white
dwarfs all have diagnostic emission or absorption features in this range.
The instrument therefore does not require a mid-infrared detector to support
the core compact-object program. Other facilities can measure an ancillary
infrared excess from cool dust when the science case requires the excess
measurement.
The wavelength span is nevertheless a demanding instrumental requirement.
Ultraviolet efficiency, optical order control, near-infrared detector
performance, and cross-channel flux calibration must be demonstrated as
separate parts of the end-to-end design. A nominal wavelength interval alone
does not establish usable sensitivity throughout the interval.

For a point source, the resolving power gives
\begin{equation}
  \Delta\lambda = \frac{\lambda}{\Rpoint},
  \qquad
  \Delta v \simeq \frac{c}{\Rpoint}.
  \label{eq:resolution}
\end{equation}
At H$\alpha$, the nominal resolution element is about $1.31$\,\AA\ in the
selectable $R=5000$ setting and about $6.56$\,\AA\ at $R=1000$. The nominal
scale does not imply that every line centroid is measured to one resolution
element.
For a sufficiently bright and isolated line, centroid precision may be much
smaller than one resolution element. The limiting factors are photon noise,
calibration stability, line blending, and the forward model of the
slitless scene.

The two resolving-power settings support different measurements. Disk lines
in cataclysmic variables often span several hundred to more than one thousand
$\kms$. Several resolution elements in the selectable $R\simeq5000$ bands
therefore sample the line wings and double peaks. Kilonova ejecta instead
produce blended structures with characteristic velocities near a tenth of the
speed of light. Detecting kilonova structure does not require $R=5000$.
Binning the native pixels increases continuum S/N at late epochs, while the
unbinned data retain narrow host-galaxy and foreground features. The proposal
must quantify the trade by convolving observed spectra with the instrument
line-spread function and injecting the convolved spectra into realistic
backgrounds.

\begin{xresultbox}{violet}
\textbf{Operational requirement.}
The spectroscopic program must preserve selectable point-source $R\simeq5000$
bands for compact objects. The wide transient baseline remains $R\simeq1000$.
Binning to a lower resolution may be used for faint continuum measurements,
but the high-resolution data must retain the line-profile information required
for velocity, profile-asymmetry, and line-ratio studies.
\end{xresultbox}

The present mission baseline does not require a high-dispersion module beyond
$R\simeq5000$. Robotic servicing could add an $R\gtrsim15000$ module after the
baseline mission has demonstrated the detector, calibration, and pointing
performance. That option is an extension path, not an assumption in the
five-year yield or time allocation below.

\section{Observables from Selectable \texorpdfstring{$R\simeq5000$}{R approximately 5000} Spectra}

The spectra provide more than a redshift or a classification label. The joint
evolution of the continuum and line profiles contains the principal physical
information. The spectral fit must report the quantities in
Table~\ref{tab:observables} together with the covariance matrix of the fitted
quantities.

The covariance is essential because slitless extraction couples neighboring
wavelength bins and contaminating traces. A profile width obtained after
instrumental deconvolution is meaningful only when the wavelength solution,
line-spread function, and source position are fitted together. Line-ratio fits
require the same joint treatment. Balmer ratios in an accretion disk depend on
optical
depth and non-local thermodynamic equilibrium as well as extinction. The
Balmer decrement must therefore enter a disk-atmosphere model instead of being
used as an isolated reddening estimator.

\begin{table}[H]
\centering
\caption{Compact-object spectral observables and the corresponding physical
interpretation.}
\label{tab:observables}
\begin{tabularx}{\textwidth}{@{}P{33mm}Y P{37mm}@{}}
\toprule
\textbf{Observable} & \textbf{Measurement} & \textbf{Physical use}\\
\midrule
Line centroid & Wavelength after wavelength calibration & Radial velocity,
redshift, orbital motion, ejecta velocity\\
FWHM and wings & Profile width after instrumental deconvolution & Disk
rotation, wind velocity, shock broadening\\
Double-peaked profile & Peak separation and asymmetry & Keplerian disk
kinematics and disk inclination constraints\\
Equivalent width & Line flux relative to local continuum & Ionization state,
accretion state, composition proxy\\
Balmer decrement & H$\alpha$/H$\beta$ and related ratios & Joint constraints
on optical depth, excitation, density, temperature, and extinction\\
He II and Bowen features & He II $\lambda4686$ and nearby blend & Hard
ionizing continuum and high-state accretion\\
Continuum slope & Multi-epoch spectral energy distribution & Temperature,
reddening, cooling, and disk contribution\\
Line evolution & Temporal and orbital-phase evolution & Reprocessing,
outflow, disk instability, and transient classification\\
\bottomrule
\end{tabularx}
\end{table}

\section{Target Selection and Reference Samples}

The compact-object survey combines a controlled reference sample,
alert-driven observations, and a blind discovery channel. The reference
sample supplies repeatable line profiles across known accretion states. The
alert channel records short-lived kilonovae and accretion episodes. A blind
extraction of the dispersed data searches for compact emission-line sources
that are absent from existing catalogs.

Catalog classifications, optical or X-ray variability, orbital information,
and predicted spectral brightness define the initial selection. The ETC then
applies the measured throughput, zodiacal background, detector configuration,
and trace-overlap loss to determine the final selection. The Ritter CV
catalog provides the initial dwarf-nova classification and subtype labels
\cite{rittercv}. LVK public alerts provide the kilonova trigger stream
\cite{lvkalerts}. A Target-of-Opportunity (ToO) observation is a rapid
follow-up observation initiated by such an alert.

\begin{table}[H]
\centering
\caption{Target-selection structure and reference observing plan. The exposure
times denote science integration across three roll angles. Slew, acquisition,
detector readout, and roll overhead are not included.}
\label{tab:target_selection}
\begin{tabularx}{\textwidth}{@{}P{31mm} P{42mm} P{38mm} Y@{}}
\toprule
\textbf{Sample} & \textbf{Selection} & \textbf{Reference exposure} &
\textbf{Campaign}\\
\midrule
Dwarf-nova intensive & Bright U Gem, SS Cyg, Z Cam, SU UMa, WZ Sge,
VW Hyi, and RU Peg with known orbital information & $3\times(300$--$600)$ s
per epoch & One year with monthly visits and 6--12 visits during each outburst\\
Dwarf-nova population & Catalogued U Gem, SU UMa, Z Cam, and WZ Sge systems
with measured or constrained brightness & $3\times(600$--$1200)$ s &
One visit every 4--6 weeks for 8--15 targets\\
KMTNet/KSP dwarf novae & Published dwarf novae discovered by the KMTNet
Supernova Program & Outburst exposures from $3\times300$ s to
$3\times1200$ s. Quiescent exposures are set by the ETC &
Regular visits for two recurrent systems and alert-driven visits for three
fainter or less frequent systems\\
AM CVn and ultracompact WD binaries & Helium emission, short orbital period,
or an established ultracompact classification & Phase bins shorter than
$0.1P_{\rm orb}$ with $1800$--$3600$ s total & Six to twelve months with
2--4 week state sampling and one phase-resolved sequence\\
UCXB benchmark & X-ray binaries with hydrogen-deficient or degenerate-donor
evidence & $2700$--$5400$ s total with phase-aware subexposures &
State-triggered visits over 6--12 months\\
WD plus NS or BH candidates & Gaia astrometric acceleration combined with
X-ray, UV, or radio evidence & $3\times(900$--$1800)$ s & Candidate confirmation
and two to four visits per year\\
Kilonova ToO & LVK BNS or NSBH alert with a visible counterpart and acceptable
solar elongation & $3\times(300$--$1200)$ s per epoch & Four epochs from 4 hr to
7 d after the alert\\
\bottomrule
\end{tabularx}
\end{table}

\begin{table}[H]
\centering
\caption{Initial compact-object candidate list. A benchmark label denotes a
well-studied system that calibrates the observing mode. A candidate label
denotes a system for which the nature of the donor or compact companion
remains part of the science case.}
\label{tab:compact_candidates}
\begin{tabularx}{\textwidth}{@{}P{34mm} P{72mm} Y@{}}
\toprule
\textbf{Class} & \textbf{Initial candidates} & \textbf{Use}\\
\midrule
Dwarf nova intensive & U Gem, SS Cyg, Z Cam, SU UMa, WZ Sge, VW Hyi, RU Peg &
Bright reference systems with complementary outburst behaviour\\
Dwarf nova population & RX And, SS Aur, ER UMa, YZ Cnc, OY Car, HT Cas,
VY Aqr, VW Vul & Spectral evolution across accretion states, inclination
effects, and population diversity\\
KMTNet/KSP dwarf novae & KSP-OT-201503a, KSP-OT-201611a, KSP-OT-201701a,
KSP-OT-201712a, KSP-OT-202104a & Ground-based discovery history combined
with regular and alert-driven space spectroscopy\\
AM CVn & AM CVn, HM Cnc, V407 Vul, ES Cet, HP Lib, CR Boo, V803 Cen, KL Dra,
GP Com, SDSS J1240$-$01, CE 315 & Helium accretion, orbital modulation, and
degenerate-donor diagnostics\\
UCXB benchmark & 4U 1820$-$30, 4U 1626$-$67, 4U 1916$-$05, 4U 0614+091,
4U 1543$-$624, 2S 0918$-$549, 4U 1850$-$087 & Hydrogen-deficient disks and
donor-composition constraints\\
WD plus NS or BH candidates & Gaia astrometric-acceleration candidates with
X-ray, UV, or radio counterparts & The list is generated from Gaia and
high-energy catalogs before the survey selection freeze\\
Kilonova reference event & GW170817 and AT2017gfo & ETC template and
multi-epoch spectral benchmark\\
Kilonova alert sample & Future LVK BNS and NSBH public alerts & ToO sample.
No fixed pre-survey object list is assumed\\
\bottomrule
\end{tabularx}
\end{table}

The candidate list does not assert that every listed system contains a white
dwarf with a neutron-star or black-hole companion. The AM CVn and UCXB groups
include systems with evidence for degenerate donors as well as systems that
may contain helium-star donors. Observable selection criteria define the WD
plus NS or BH sample because the available catalogs do not yet provide a
verified object list.

The shortest-period systems require a separate temporal design. HM~Cnc and
V407~Vul show modulations near 321 and 569 s. The UCXB 4U~1820$-$30 has an
orbital period near 685 s \cite{barros2007,stella1987}. A 600 s exposure
averages over a substantial fraction of an orbit and can smear the velocity
structure that motivates phase-resolved spectroscopy. The total
integration in Table~\ref{tab:target_selection} must be divided into short
subexposures for bright targets. Fainter systems provide a state-averaged
spectrum, but the resulting product must not be described as orbital
tomography.

The reference table does not constitute a mission time allocation. The annual
time requirement is formulated as
\begin{equation}
 T_{\rm program} =
 \sum_k N_{{\rm target},k}N_{{\rm epoch},k}
 \left(t_{{\rm int},k}+t_{{\rm roll},k}+t_{{\rm read},k}\right)
 + T_{\rm slew}+T_{\rm ToO},
 \label{eq:time_budget}
\end{equation}
where the index $k$ denotes a source class. For bright dwarf novae, the roll
and slew terms can exceed the detector integration. The final proposal must
evaluate Equation~\ref{eq:time_budget} with the spacecraft attitude model. The
resulting time requirements and predicted scientific return must rank the
source classes before a five-year yield is assigned.

\subsection{KMTNet/KSP dwarf-nova legacy sample}

The Korea Microlensing Telescope Network (KMTNet) operates three 1.6-m
wide-field telescopes in Chile, South Africa, and Australia
\cite{kim2016kmtnet}. The longitude distribution supports nearly continuous
ground-based photometry when weather and seasonal visibility permit. The
KMTNet Supernova Program (KSP) has published five dwarf-nova discoveries with
measured light-curve histories or spectroscopic classifications
\cite{brown2018ksp,lee2019ksp,lee2022ksp,lee2024ksp,kim2026ksp}.

The five discoveries form a standing KMTNet/KSP legacy sample for the
spacecraft. The standing sample does not require a common observing interval.
Quiescent brightness differs by several magnitudes, and recurrence time spans
more than an order of magnitude. Uniform sampling would require excessive
quiescent exposure time for KSP-OT-202104a and would undersample the 6.6-day
activity of KSP-OT-201712a. Table~\ref{tab:ksp_dwarf_novae} assigns an initial
cadence tier from the published behaviour. The ETC and spacecraft visibility
model must determine the final exposure and visit count.

\begin{table}[H]
\centering
\small
\caption{Published KMTNet/KSP dwarf novae proposed as a standing monitoring
sample. Coordinates use the J2000 reference frame. The cadence tiers are
initial observing assumptions rather than guaranteed mission allocations.}
\label{tab:ksp_dwarf_novae}
\begin{tabularx}{\textwidth}{@{}P{28mm} P{31mm} P{57mm} Y@{}}
\toprule
\textbf{Source} & \textbf{J2000 position} & \textbf{Published behaviour} &
\textbf{Proposed monitoring}\\
\midrule
KSP-OT-201503a \cite{brown2018ksp} &
$09^{\rm h}23^{\rm m}41.40^{\rm s}$\newline
$-21^\circ58^\prime09.5^{\prime\prime}$ &
U Gem-type candidate with $V\simeq17.3$ at maximum and $V\simeq22.6$ in
quiescence. The 2015 event lasted about 17 d. At least three events occurred
from 2011 to 2015. &
Tier 2. Observe every 3 to 4 weeks and begin daily spectroscopy after a
KMTNet rise alert. Follow the Balmer profile through the outside-in heating
front.\\
KSP-OT-201611a \cite{lee2019ksp} &
$06^{\rm h}42^{\rm m}02.1^{\rm s}$\newline
$-26^\circ00^\prime21.6^{\prime\prime}$ &
SU UMa or U Gem-type system with $V\simeq23.45$ in quiescence. A long
outburst and a short outburst were separated by about 91 d. The large
Galactic height makes the source a Population II candidate. &
Tier 2. Observe every 2 to 4 weeks within a seasonal campaign and increase
the cadence after an optical alert. Compare disk-line evolution between long
and short events.\\
KSP-OT-201701a \cite{lee2022ksp} &
$06^{\rm h}39^{\rm m}23.2^{\rm s}$\newline
$-26^\circ37^\prime18.8^{\prime\prime}$ &
Three superoutbursts and six normal outbursts were measured over 2.6 yr.
The normal cycle is about 76 d. The orbital period is
$51.91\pm2.50$ min. Strong double-peaked H~I and weak He~I emission are
present. &
Tier 1. Observe every 7 to 14 d and use daily visits during an outburst.
Measure peak separation, He~I/H$\alpha$ and the transition toward a
helium cataclysmic variable.\\
KSP-OT-201712a \cite{lee2024ksp} &
$07^{\rm h}24^{\rm m}07.70^{\rm s}$\newline
$-26^\circ19^\prime54.5^{\prime\prime}$ &
ER UMa-type system below the period minimum with a
$58.75\pm0.02$ min orbit. Outbursts recur every 6.6 d during standstills and
about every 15 d outside the standstills. The spectrum contains double-peaked
H$\alpha$ and weak He~I. &
Tier 1. Observe every 2 to 3 d during focused 30 to 60 d windows. Resolve
line-profile evolution across standstills and determine the response to the
high mass-transfer state.\\
KSP-OT-202104a \cite{kim2026ksp} &
$08^{\rm h}14^{\rm m}20.14^{\rm s}$\newline
$-26^\circ02^\prime37.21^{\prime\prime}$ &
Type D WZ Sge system with an amplitude near 8 mag and a 28.5 d outburst.
The superhump period is about 71.7 min. The measured brightness spans
$V\simeq24.05$ in quiescence to $V\simeq16.01$ at maximum. &
Tier 3. Retain a low-cadence state check only when the ETC supports the
quiescent exposure. A KMTNet alert starts intensive spectroscopy within
12 hr and daily visits follow the early outburst.\\
\bottomrule
\end{tabularx}
\end{table}

KSP-OT-201712a and KSP-OT-201701a receive the regular allocation because the
recurrence times allow the mission to sample multiple disk states.
KSP-OT-201503a and KSP-OT-201611a retain a sparse baseline that establishes
the pre-outburst spectrum. KSP-OT-202104a is dominated by the alert mode
because the quiescent magnitude requires a long exposure, while the outburst
remains bright for several weeks.

The $R=5000$ sequence links the KMTNet light curve to velocity-resolved
Balmer and helium profiles. The sequence measures disk peak separation,
profile asymmetry, line ratios, and the transition between quiescent emission and
outburst absorption. The KMTNet/KSP objects do not replace the bright
reference systems in Table~\ref{tab:compact_candidates}. The bright systems
measure instrumental repeatability while the KMTNet/KSP sample tests the
scientific reach at faint flux and unusual evolutionary state.

\subsection{KMTNet transient alerts and near-infrared spectroscopy}

KMTNet transient observations can trigger the compact-object program. Each
notice supplies a discovery time, a sky position, and an optical light curve
for ranking the target before the spacecraft begins a spectroscopic visit. The
notice does not replace the space-based spectrum. The KMTNet observation
identifies the transient state, while the space-based observation measures the
spectrum of the transient state.

The standalone appendix does not yet define the KMTNet/KSP alert performance.
The main proposal must specify the adopted field of view, filter set,
single-visit depth, revisit interval, astrometric uncertainty, classification
latency, and annual duty cycle. The survey parameters determine whether KMTNet
discovers an outburst before the rise is complete and whether a kilonova
candidate is localized well enough for immediate spectroscopy. An alert
mechanism without measured latency and sky coverage does not demonstrate
operational feasibility.

Near-infrared spectroscopy is especially valuable for transient targets. The
reddest ejecta component of a kilonova may become dominant after the first
optical observations. Dust extinction is also weaker in the near-infrared than
in the optical. The $1.0$--$1.5\micron$ portion of the adopted band can
therefore measure a red component that is attenuated or missed at shorter
wavelengths.
Space-based operation removes telluric absorption and provides a stable
background for repeated line-flux and continuum measurements. The
near-infrared range also contains Paschen and helium diagnostics that
complement the Balmer lines used for dwarf novae and accretion binaries.

The KMTNet/KSP notice supplies the position, discovery time, photometric
state, and classification probability. The mission scheduler then evaluates
solar elongation, visibility duration, slew cost, and the predicted spectral S/N.
A bright target receives an initial $3\times300$ s sequence. Later visits use
$3\times600$--$1200$ s when the source has faded. The direct-image scene
determines the number of roll angles. Two orientations can be sufficient for
an isolated target, while a crowded field can require a third orientation.
The released product combines the KMTNet light curve with the calibrated
optical and near-infrared spectra.

KMTNet alerts identify dwarf-nova outbursts, transitions between accretion
states, and optical counterparts of compact mergers. The space-based near-infrared
spectrum supplies information that a ground-based optical notice does not
provide. The proposed program should reserve ToO time for the alert stream in
addition to the standing KMTNet/KSP dwarf-nova sample.

The priority decision should use a reproducible score based on the probability
of the source class, the probability of spacecraft visibility, the predicted
S/N, and the remaining lifetime of the transient state. A kilonova candidate
with a rapidly fading near-infrared prediction has a different time cost from
a dwarf nova that remains near maximum for several days. The alert service
must preserve the selection probabilities and the origin of every datum rather
than distribute only a binary target flag.

The analysis must distinguish Doppler broadening from morphological broadening
along the dispersion direction. The small angular size of a compact object
minimizes morphological broadening. The forward model must still propagate the
point-spread function, trace curvature, detector sampling, and overlap from
unrelated field sources.

The scene model must be tested with a transient that brightens or fades during
the sequence. A static contamination template can assign variable flux to the
wrong trace when the target brightens between the direct image and the
dispersed exposure. The simulation should include the temporal mismatch and
report the recovered line flux as a function of target contrast, source
separation, and roll angle.

\section{Gravitational-Wave Counterparts}

\subsection{Binary neutron-star and neutron-star--black-hole mergers}

Binary neutron-star mergers demonstrate the need for rapid space-based
spectroscopy. The gravitational-wave signal
constrains the binary masses, spins, distance, inclination, and tidal
deformability. The electromagnetic counterpart constrains the ejecta, viewing
angle, host galaxy, and radioactive heating. GW170817
demonstrated that the optical and near-infrared emission evolves on a
timescale of days and that different ejecta components may dominate at
different wavelengths \cite{abbott2017,arcavi2017,tanaka2017}.

The compact-object program does not require a mid-infrared channel for the
central measurement. Repeated imaging and spectroscopy follow the early blue
component and the later redder component from the near-ultraviolet through the
near-infrared. The spectral sequence constrains
temperature evolution, expansion velocity, opacity, and the relative
contribution of high- and low-electron-fraction ejecta. The identification of
strontium in the spectrum of GW170817 linked the kilonova directly to
nucleosynthesis rather than treating the event as a purely
photometric transient \cite{watson2019}.

The red wavelength limit defines the fraction of the spectral evolution that
the mission measures independently. A lanthanide-rich component may peak
beyond $1.5\,\micron$ after the first few days. The adopted instrument measures
the blue component, the transition into the near-infrared, and the
short-wavelength side of the red component. Measuring the full bolometric
luminosity or the late lanthanide-rich ejecta requires deeper infrared
observations from another facility. Radiative-transfer templates must quantify
the information lost at the $1.5\,\micron$ boundary. The proposal cannot claim
an independent measurement of the full red component.

The large ejecta velocity shifts the principal requirement from native
resolving power to continuum sensitivity and wavelength coverage. Kilonova
features with velocities near $0.1c$ are strongly blended
\cite{pian2017,chornock2017}. Continuum S/N, wavelength coverage, and response
time primarily determine detectability. The $R=5000$ pixels
should be combined into lower-resolution spectral bins for ejecta inference.
The native data remain useful for a narrow host line, an intervening absorber,
or a contaminating field source. The observing mode must return both products
from one calibrated likelihood.

Each event requires a joint fit. The observables entering the fit are
formulated as
\begin{equation}
  \boldsymbol{D}_{\rm KN} =
  \left\{F_\lambda(t),\,F_b(t),\,z_{\rm host},\,
  p(d_L,\iota\mid D_{\rm GW})\right\},
  \label{eq:kilonova_parameters}
\end{equation}
where $F_\lambda(t)$ is the spectral sequence, $F_b(t)$ denotes the
multi-band light curves, and $z_{\rm host}$ is the host redshift. The final
term is the gravitational-wave posterior for luminosity distance $d_L$ and
inclination $\iota$. A radiative-transfer model maps the data vector to the masses,
velocities, electron-fraction distributions, and geometry of the dynamical
and wind ejecta. Opacity is wavelength dependent and evolves with ionization.
The opacity should not be treated as a single independent number without a
specified model. The telescope constrains the electromagnetic likelihood. The
nuclear equation of state enters only through joint inference with the
gravitational-wave posterior and merger simulations.

\begin{figure}[tp]
\centering
\includegraphics[width=0.95\textwidth,height=0.62\textheight]{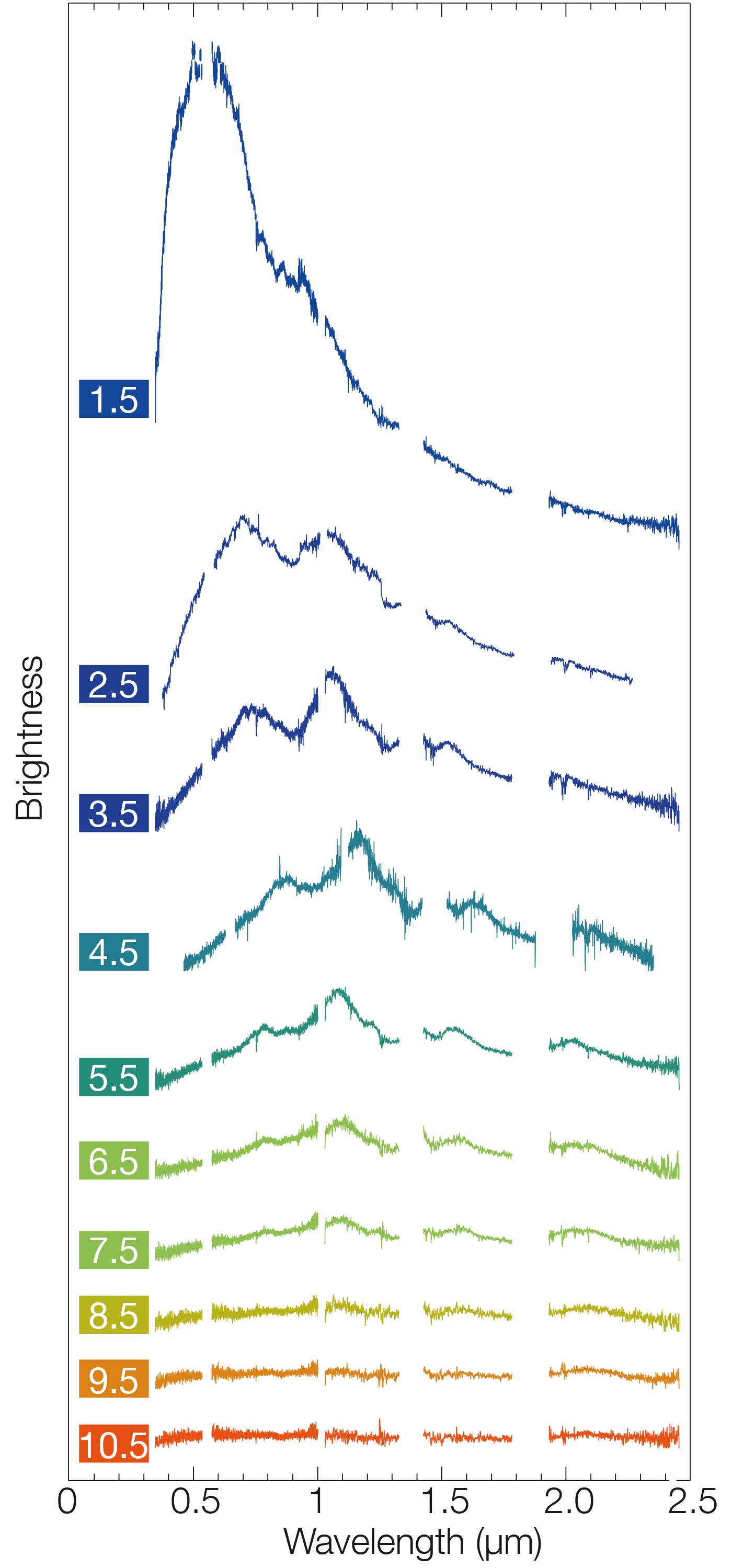}
\caption{Actual observed optical and near-infrared spectra of the kilonova
AT~2017gfo produced by GW170817, obtained with the ESO VLT/X-shooter
instrument over twelve days. The sequence becomes progressively redder and
develops broad, evolving absorption structure, corresponding to the kilonova
rows in Table~\ref{tab:dwarf_lines}. The published reference montage extends
to about $2.5\,\micron$. The proposed baseline uses the $0.2$--$1.5\,\micron$
portion and does not require a mid-infrared mode. Reproduced from the ESO
public image archive \cite{eso1733j}. The physical interpretation is
discussed by Tanaka et al. (2017) and Watson et al. (2019)
\cite{tanaka2017,watson2019}. Courtesy of ESO.}
\label{fig:kilonova_observed}
\end{figure}

\subsection{ToO sequence and response requirement}

The proposed observing sequence prioritizes the short-lived early signal.
Table~\ref{tab:too} gives a representative sequence.
The exact number of visits must be recomputed after the spacecraft slew,
solar-avoidance, and alert-localization simulations are complete.

The response clock begins at the external alert time and ends at the midpoint
of the first science exposure. Spacecraft slew time alone understates the
latency because counterpart identification, command authorization, guide-star
acquisition, and detector configuration occur before useful photons are
recorded. The first visit should use the minimum number of roll angles
supported by the direct-image scene. A mandatory
three-roll sequence may consume the interval in which the ultraviolet and blue
components evolve most rapidly in flux and color.

\begin{table}[tbp]
\centering
\caption{Reference compact-object ToO sequence.}
\label{tab:too}
\begin{tabularx}{\textwidth}{@{}C{22mm} C{29mm} Y Y@{}}
\toprule
\textbf{Elapsed time} & \textbf{Mode} & \textbf{Primary information} & \textbf{Requirement}\\
\midrule
$\simeq2$ hr goal & Imaging in several bands & Localization, color, counterpart ranking &
Rapid alert ingestion, slew, and acquisition\\
$2$--$12$ hr & Selectable $R\simeq5000$ spectrum & Early line and continuum state &
Stable wavelength calibration and sufficient S/N per resolution element\\
$1$--$3$ d & Repeated spectroscopy & Ejecta velocity and ionization evolution &
Repeatable trace and flux calibration\\
$4$--$14$ d & Imaging plus spectroscopy & Cooling, recombination, and red component &
Low background and flexible queue scheduling\\
$>14$ d & Host and late-time imaging & Host redshift and transient subtraction &
Accurate reference imaging\\
\bottomrule
\end{tabularx}
\end{table}

The planning goal is approximately two hours from alert to the first useful
exposure. A fallback response below four hours preserves the early kilonova
sequence. The mission-level requirement extends beyond a fast slew. Before
the first exposure, the observatory must validate the alert, calculate visibility,
update the queue, acquire the target, assess data quality, and disseminate the
result. A rapid spacecraft slew cannot recover the early kilonova signal when
ground-segment processing delays the first exposure.

The expected yield must be stated separately from the response goal. For an
observing interval $T$, the number of useful spectral sequences is formulated
as
\begin{equation}
 N_{\rm use}=T\int R_{\rm merger}(D)
 f_{\rm GW}(D)f_{\rm loc}(D)f_{\rm vis}
 f_{\rm id}f_{\rm S/N}(D)\,dV ,
 \label{eq:kilonova_yield}
\end{equation}
where $R_{\rm merger}$ is the merger rate density. The factors quantify
gravitational-wave detection, useful localization, spacecraft visibility,
counterpart identification, and adequate spectral S/N. Every factor depends
on the observing network, and several factors depend on distance. The mission
cannot replace the integral with the number of public alerts. A credible
five-year estimate must quote the median yield and the uncertainty from the
merger-rate posterior.

\begin{figure}[H]
\centering
\begin{tikzpicture}[x=1cm,y=1cm,font=\sffamily\footnotesize]
  \draw[->,very thick,color=violet] (0,0) -- (13.5,0);
  \foreach \x/\t in {0/{GW alert},2/{visibility},4/{slew},6/{first color},8.5/{spectrum},11/{repeat},13/{host}} {
    \fill[violet] (\x,0) circle (2.2pt);
    \node[align=center,anchor=north] at (\x,-0.12) {\t};
  }
  \draw[<->,color=rose,line width=1.2pt] (0,0.7)--(4,0.7)
    node[midway,above] {$\simeq2$ hr planning goal};
  \draw[<->,color=teal,line width=1.2pt] (4,1.35)--(8.5,1.35)
    node[midway,above] {early evolution};
  \draw[<->,color=gold,line width=1.2pt] (8.5,2.0)--(13,2.0)
    node[midway,above] {velocity and cooling sequence};
\end{tikzpicture}
\caption{The compact-object ToO sequence. The science requirement is a timed
sequence of measurements rather than a single rapid exposure.}
\label{fig:too_timeline}
\end{figure}
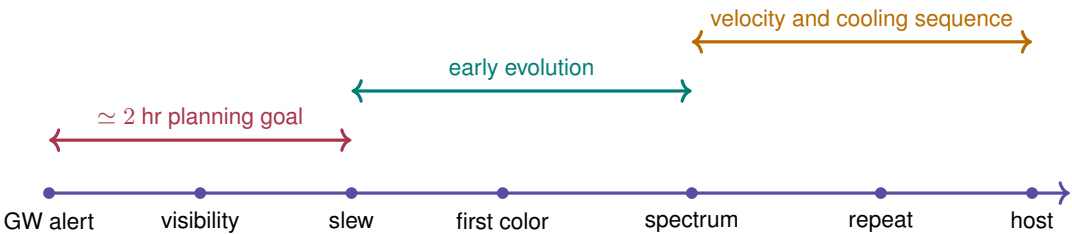

\section{Dwarf Novae and Cataclysmic Variables}

Dwarf novae provide repeatable targets for a robotic $R=5000$ slitless
spectrograph. Dwarf novae are unresolved and recurrently variable, with rich
emission-line and absorption-line spectra. The standard disk-instability model
attributes dwarf-nova outbursts to a thermal-viscous instability in the
accretion disk of a close binary
\cite{warner1995,lasota2001}. The observed spectrum records the transition
between the quiescent and outburst states. Repeated spectra tie the disk
temperature and velocity field to the light curve instead of providing only a
static classification.

\begin{figure}[H]
\centering
\includegraphics[width=0.94\textwidth]{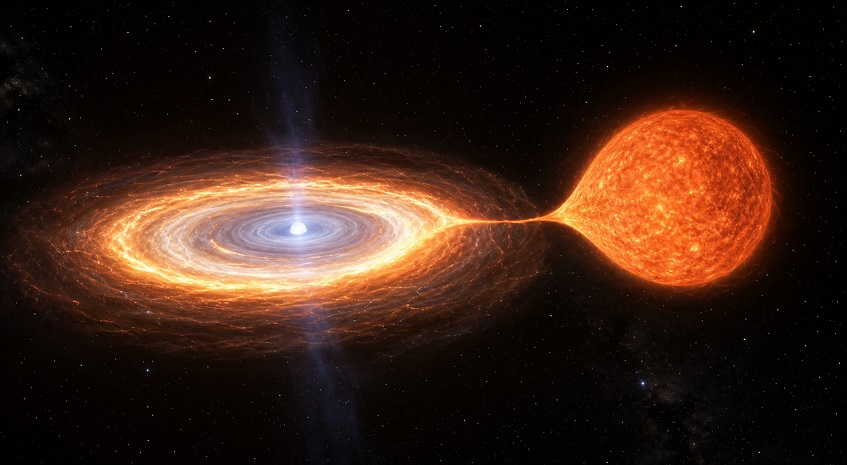}
\caption{Artistic reconstruction of a typical dwarf-nova binary during a disk
outburst. A low-mass late-type donor transfers gas through the inner Lagrange
point to a white dwarf. The bright annulus represents a heating front moving
through the accretion disk. The image is a physical illustration rather than
an observed image.}
\label{fig:dwarf_nova_concept}
\end{figure}

\subsection{Where the outburst occurs}

The dwarf-nova outburst occurs in the accretion disk. The white dwarf remains
the compact central accretor, but the sudden optical brightening is not a
thermonuclear explosion on the white-dwarf surface. The distinction separates
a dwarf nova from a classical nova.

A typical system contains a white dwarf and a Roche-lobe-filling late-type
main-sequence donor. The donor loses gas through the inner Lagrange point.
The transferred gas carries angular momentum and forms a disk around the
white dwarf. The mass-transfer rate can remain below the threshold required
for a stable hot state.

The disk instability is a limit cycle rather than an explosion. Mass
accumulates while hydrogen is mostly neutral and angular-momentum transport is
inefficient. An annulus becomes thermally unstable when the surface density
crosses the critical value at partial hydrogen ionization. The
increased opacity and ionization move the annulus to the hot branch. A heating
front then travels through the disk and raises the inward mass flux. The
optical and ultraviolet continuum brightens as a larger disk area enters the
hot state. Depletion eventually permits a cooling front to form. The cooling
front returns the disk to the low-ionization branch and begins the next
quiescent interval.

Local thermal equilibrium produces the familiar S-shaped relation between
surface density and effective temperature \cite{lasota2001}. The lower and
upper branches are stable over different surface-density intervals. The
middle branch near $6500$--$10{,}000$ K is unstable. The standard
parameterization uses a larger effective viscosity in the hot state than in
the cool state. The measured rise and decay do not determine the hot-state and
cool-state viscosities directly. Interpreting the light curve requires a disk
model in which the thermal and viscous times are
\begin{equation}
 t_{\rm th}\sim \frac{1}{\alpha\Omega_{\rm K}},
 \qquad
 t_{\rm visc}\sim \frac{1}{\alpha\Omega_{\rm K}}
 \left(\frac{R}{H}\right)^2 ,
 \label{eq:disk_times}
\end{equation}
where $\alpha$ is the effective viscosity parameter, $\Omega_{\rm K}$ is the
Keplerian angular frequency, and $H/R$ is the disk aspect ratio. A heating
front shifts the local spectrum from the cool branch to the hot branch on a
timescale closer to $t_{\rm th}$. The
redistribution of disk mass follows the longer viscous time. Spectra obtained
only at maximum light do not separate heating-front propagation from viscous
mass redistribution.

An outside-in outburst begins in the outer disk when mass transfer and surface
density are high there. An inside-out event begins closer to the white dwarf
after the inner disk reaches the critical density. The propagation direction
sets the order in which the continuum, line wings, and peak separation
respond. A combined light curve and spectral sequence therefore tests the
front direction while constraining disk radius and mass-transfer state.

The optical rise does not indicate an explosion of the white dwarf. The heated
disk produces most of the optical luminosity, with additional reprocessed
emission from the donor and the disk surface. Dwarf-nova spectra should
consequently change from emission-line-dominated quiescence to a brighter
continuum with broad absorption or mixed emission and absorption during
outburst.
The U~Gem spectrum in Figure~\ref{fig:ugem_observed} provides a direct
observed example of the state dependence.

\subsection{Spectral features in the adopted wavelength range}

The optical and near-infrared range contains the principal spectroscopic
diagnostics of dwarf-nova accretion states.

\begin{table}[H]
\centering
\caption{Compact-object diagnostics available without a mid-infrared channel.
The dwarf-nova rows correspond to the observed U~Gem spectrum in
Figure~\ref{fig:ugem_observed}. The kilonova row corresponds to the observed
AT~2017gfo sequence in Figure~\ref{fig:kilonova_observed}.}
\label{tab:dwarf_lines}
\begin{tabularx}{\textwidth}{@{}P{28mm} P{37mm} Y@{}}
\toprule
\textbf{Feature} & \textbf{Typical use} & \textbf{Caveat}\\
\midrule
Balmer series & Disk temperature, optical depth, velocity field, extinction &
Profiles may switch from absorption to emission during outburst\\
H$\alpha$ & Strong time-series line and radial-velocity tracer &
May be contaminated by winds or the irradiated donor\\
He I lines & Ionization and disk-state indicator &
Several transitions are blended at low S/N\\
He II $\lambda4686$ & Hard ionizing component and high-state accretion &
Requires adequate blue sensitivity and calibration\\
Bowen blend near $4640$\,\AA & Reprocessed high-energy radiation and irradiated gas &
Blended components require forward profile fitting\\
Ca II infrared triplet & Cool dense gas and chromospheric or disk contribution &
Useful mainly for selected systems\\
Paschen lines & Near-infrared disk and wind diagnostics &
Telluric absorption is absent in space, which improves repeatability\\
Kilonova broad absorption complexes & Expansion velocity, temperature,
opacity, and evolution of lanthanide-rich and lanthanide-poor ejecta &
Features are broad and blended. Interpretation requires radiative-transfer
models and the external gravitational-wave posterior\\
Kilonova continuum sequence & Cooling, recombination, and radioactive-heating
timescale &
The early spectrum must be obtained quickly because the color evolves over
days\\
\bottomrule
\end{tabularx}
\end{table}

\begin{figure}[H]
\centering
\includegraphics[width=0.83\textwidth]{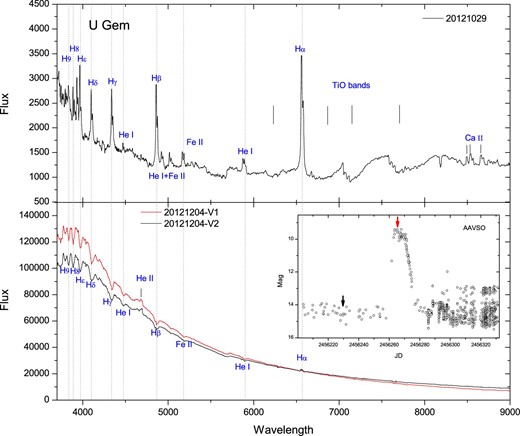}
\caption{Actual observed spectra of the dwarf nova U~Gem from LAMOST. The
upper spectrum was obtained in quiescence and the lower spectra near the
2012 December outburst peak. The published labels identify Balmer lines,
He~I, He~II, Fe~II, TiO bands, and the Ca~II infrared triplet. The labelled features
are the observed examples represented by the dwarf-nova rows in
Table~\ref{tab:dwarf_lines}. Reproduced from Zhi et al. (2020), Figure 1
\cite{zhi2020}. Courtesy of Zhi et al. (2020).}
\label{fig:ugem_observed}
\end{figure}

Figure~\ref{fig:ugem_observed} shows why the table must distinguish the
quiescent and outburst states. In quiescence, U~Gem has strong Balmer and
helium emission together with Ca~II emission and donor-star absorption bands.
Near outburst maximum, the continuum becomes much brighter and bluer, while
broad absorption features dominate and He~II may appear in emission. The
observed transition demonstrates that a spectral time series follows the
disk-instability cycle rather than merely classifying the object.

At $R=5000$, the spectrum resolves broad disk-line structure while retaining
sufficient continuum sensitivity for the reference survey. The nominal
velocity scale is about $60\,\kms$. A high-S/N centroid or a phase-folded line
profile can yield a substantially smaller statistical uncertainty. The line
profile as a function of orbital phase and accretion state is the principal
physical data product.

\subsection{Research enabled by the spectra}

The double-peaked profile provides the most direct observable of disk
kinematics.
For an approximately Keplerian emitting disk, the peak separation scales as
\begin{equation}
 \Delta v_{\rm peak}\simeq
 2\left(\frac{GM_{\rm WD}}{R_{\rm em}}\right)^{1/2}\sin i ,
 \label{eq:double_peak}
\end{equation}
where $M_{\rm WD}$ is the white-dwarf mass, $R_{\rm em}$ is the characteristic
outer line-emitting radius, and $i$ is the binary inclination
\cite{horne1986}. The high-velocity wings arise at smaller radii. A shift in
peak separation does not by itself prove that the physical disk edge moved
because the emissivity distribution and optical depth also evolve during an
outburst. Multi-line fitting and an inclination prior are required.

The line and continuum sequences constrain complementary aspects of the
accretion state. Equivalent widths fall when a bright continuum dilutes an
emission line even if the line luminosity remains constant. Absolute
flux-calibrated line measurements are therefore required in addition to
equivalent widths. Blue-shifted absorption and asymmetric wings identify a
wind only when the profile structure is reproduced at more than one roll angle
and is inconsistent with a contaminating trace. Balmer, helium, and Paschen
profiles then test whether the outflow strengthens or weakens as disk
ionization rises.

Phase-resolved centroids measure the velocity of an irradiated donor or a
localized disk component. An emission-line velocity does not automatically
trace the white dwarf. The inferred semi-amplitude must be combined with
eclipse geometry, photometric ephemerides, and a model for the emitting site.
Applying the phase-resolved analysis across rising, maximum, and declining
states tests whether the disk-instability sequence predicts the observed front propagation
and line-forming radius. A population sample then relates the state sequence
to orbital period, recurrence time, and secular mass-transfer rate.

\begin{figure}[tp]
\centering
\includegraphics[width=0.78\textwidth]{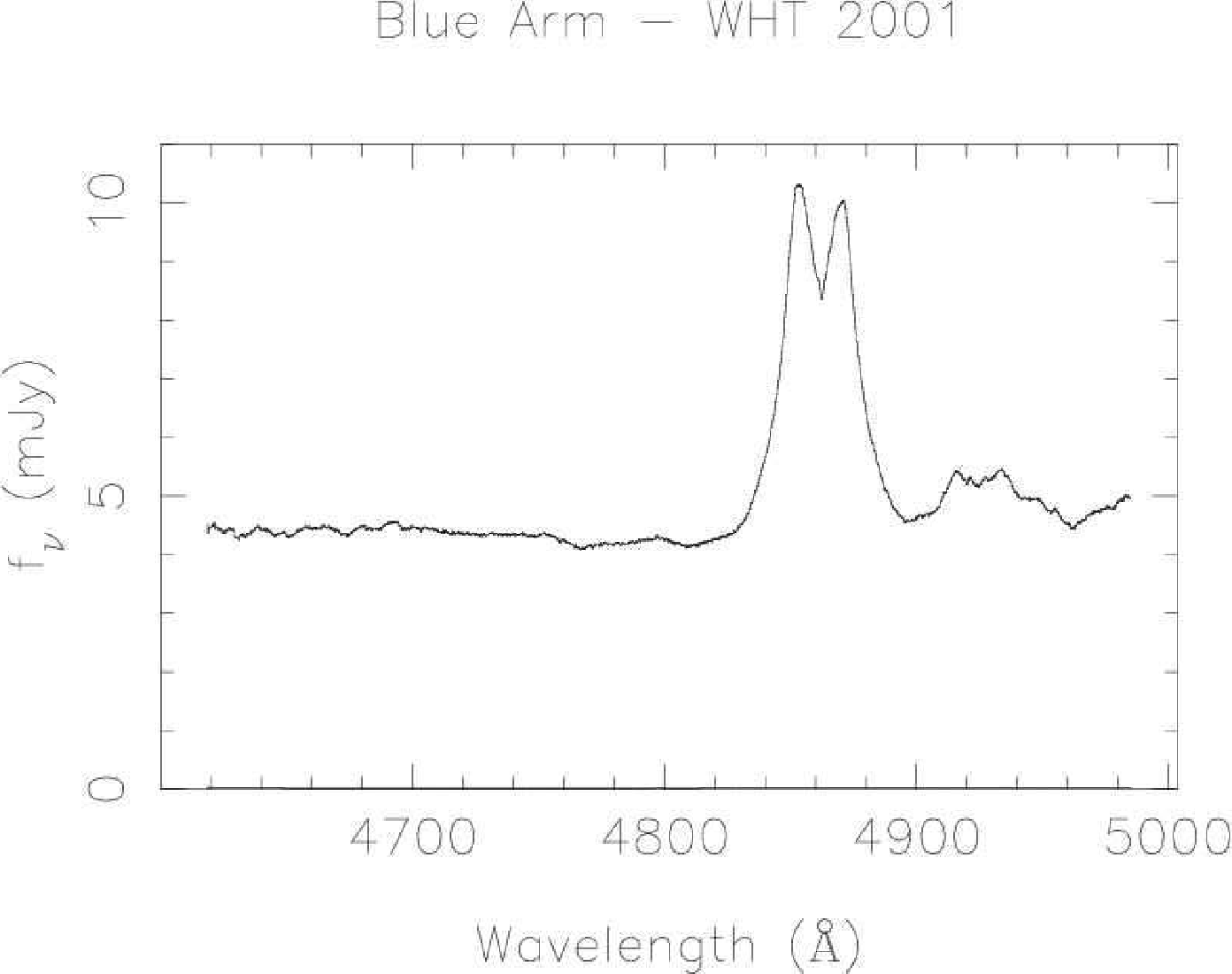}
\caption{Observed high-dispersion H$\beta$ profile of the dwarf nova U~Gem
in quiescence. The two maxima on either side of 4861\,\AA\ are the
approaching and receding sides of the rotating accretion disk. The central
depression is resolved directly in the mean WHT/ISIS spectrum. The
instrumental FWHM was 0.42\,\AA, which corresponds to
$R\simeq11{,}600$ and about $26\,\kms$ at H$\beta$. The proposed
$R=5000$ mode would sample the measured peak separation while retaining
survey efficiency. Cropped from Figure~1 of Unda-Sanzana et al. (2006)
\cite{undasanzana2006}. Courtesy of Unda-Sanzana et al. (2006).}
\label{fig:dwarf_profile}
\end{figure}

\subsection{Cadence for dwarf novae}

One spectrum per year cannot resolve dwarf-nova state evolution. Sparse
imaging first determines the quiescent level and
identifies the outburst onset. Triggered spectroscopy then samples the rise,
maximum, and decline. A selected bright subset receives a dense sequence over
at least one binary orbit. The three temporal scales measure recurrence,
thermal-state evolution, and orbital kinematics, respectively.

Orbital tomography requires approximately uniform phase coverage. To obtain
$N_\phi$ independent bins, an individual integration should satisfy
\begin{equation}
 t_{\rm exp}\lesssim \frac{P_{\rm orb}}{N_\phi}.
 \label{eq:phase_smear}
\end{equation}
A reference value of $N_\phi=10$ limits phase smearing to one tenth of an
orbit. Detector read noise and readout overhead determine whether a given
source satisfies the exposure-time condition. The observing plan should report the
number of usable phase bins after occultations, rolls, and rejected frames.
The monthly and weekly cadences in Table~\ref{tab:target_selection} describe
state sampling. The monthly and weekly cadences do not replace the dense
sequence needed for tomography.

The direct and dispersed images permit a joint fit of the target and
neighboring field objects across the full sample. A nearby star can produce an
overlapping trace. A static continuum template can miss a dwarf nova in a
crowded field. The reduction must
fit the direct image, all available roll angles, the spectral trace,
and the time-dependent source flux in one forward model.

\section{White Dwarfs and Dense-Matter Laboratories}

White dwarfs connect compact-object observations to stellar evolution,
crystallization, and the fate of planetary systems. Comparison with
stellar-structure models allows white-dwarf pulsations to probe internal
chemical stratification, total mass, rotation, and magnetic-field properties
\cite{corsico2019}. The imaging channel provides the long uninterrupted light
curves needed for asteroseismology. The $R=5000$ spectrum provides effective
temperature, surface gravity, atmospheric composition, radial velocity, and
metal-pollution diagnostics.

The useful pulsators occupy restricted temperature intervals. Hydrogen
atmosphere ZZ~Ceti stars, helium atmosphere V777~Her stars, and hotter
GW~Vir stars require different atmosphere grids and span different ranges of
periods. Spectroscopy first places a target relative to the appropriate
instability strip. The imaging sequence then measures frequencies for
comparison with the eigenmodes of a model that uses the measured atmospheric
boundary condition.

Cadence and duration control different parts of the experiment. A sampling
interval $\Delta t$ gives a Nyquist frequency
\begin{equation}
 \nu_{\rm Nyq}=\frac{1}{2\Delta t},
 \qquad
 \delta\nu_{\rm run}\simeq\frac{1}{T_{\rm run}},
 \label{eq:wd_sampling}
\end{equation}
where $T_{\rm run}$ is the uninterrupted duration. Sub-minute imaging
samples short pulsations while a run of many hours resolves nearby modes and
reduces daily aliasing. The proposal must state the photometric precision per
sample and the expected mode-amplitude threshold. A cadence value without the
photometric precision and mode-amplitude threshold does not establish an
asteroseismic yield.

For a polluted white dwarf, the optical spectrum identifies or constrains
elements such as Ca, Mg, Fe, and Si. The interpretation requires diffusion and
accretion models. The telescope does not directly measure the composition of
the disrupted parent body. The spectrum measures atmospheric abundances. A
diffusion and accretion model then infers the parent-body composition.

The atmospheric data product consists of a calibrated spectrum, effective
temperature, surface gravity, atmospheric class, line abundances, and radial
velocity. The interior data product consists of pulsation frequencies, amplitudes,
and temporal evolution. The separation avoids attributing an interior
measurement to spectroscopy alone. The time-series photometry and the
spectrum are complementary parts of the same white-dwarf experiment.

The resolving power also sets a limit on the abundance claim. Strong Ca
features and broad hydrogen or helium profiles are measurable at $R=5000$ for
sufficiently bright stars. Weak blends and detailed abundance patterns often
require higher resolution from a ground-based spectrograph. The space program
should be presented as a homogeneous discovery and atmospheric-characterization
survey followed by high-resolution work on the most informative polluted
systems.

\section{Black-Hole and Neutron-Star Accretion Transients}

Compact accretors that are not detected through a gravitational-wave alert
also produce useful optical and near-infrared time series. Stellar-mass black
hole binaries and neutron-star X-ray binaries show rapid evolution in
continuum slope, emission lines, and line asymmetry. X-ray and gamma-ray
facilities provide the high-energy trigger while the proposed telescope
measures the lower-energy response. Tidal-disruption events are not included
in the baseline sample because the extended hosts and nuclear positions
create a different slitless-extraction problem.

The optical and near-infrared continuum separates a thermal disk component
from synchrotron emission only when the spectral sequence is fitted jointly with
the X-ray state. He~II and the Bowen blend trace irradiation by the
high-energy source. The velocity curves may reveal an irradiated donor in
selected systems, but disk and wind components may shift the measured
centroid. Blue-shifted absorption and evolving line wings provide evidence for
an outflow when the wavelength calibration and comparison epochs exclude a
trace artifact. Repeated imaging adds periodic or quasi-periodic modulation
and ties each spectrum to the accretion state.

The trigger sample should be restricted to systems for which the predicted
optical or near-infrared flux reaches the line-S/N requirement. An X-ray alert
alone does not guarantee a useful spectrum. The target decision requires an
updated counterpart position, extinction estimate, current optical magnitude,
and the predicted lifetime of the state. Bright benchmark binaries
calibrate the relation between the high-energy state and the lower-energy
spectrum. Fainter events should enter only when the ETC predicts a measurable
continuum or a specified line flux.

Joint observations distinguish a disk-dominated flare from a jet-dominated
event only when the spectra are fitted with multi-band light curves and
external high-energy measurements. The program therefore coordinates the
space-based spectra with ground-based photometry and high-energy measurements.
The compact-object program does not operate as a standalone optical
classification survey.

\section{Science-to-Requirement Feasibility}

\begin{table}[H]
\centering
\caption{Reference requirements for the compact-object program. The values are
planning values to be verified by an end-to-end ETC and scene simulation.}
\label{tab:requirements}
\begin{tabularx}{\textwidth}{@{}P{43mm} P{30mm} Y@{}}
\toprule
\textbf{Requirement} & \textbf{Reference value} & \textbf{Reason}\\
\midrule
Wavelength range & $0.2$--$1.5\micron$ baseline. $2.70\micron$ throughput
threshold and $3.0\micron$ operational goal & Covers ultraviolet/optical
transient continuum, Balmer and helium lines, and selected NIR transitions.
The $2.5\micron$ band is the formal engineering off-ramp. Each detector
channel requires an independent sensitivity budget\\
Point-source resolving power & $R\simeq1000$ baseline. Selectable
$R\simeq5000$ bands & The baseline supports broad transient spectra. The
selectable bands resolve approximately $60\,\kms$ velocity structure in
accretion-disk profiles. Broad transient features are binned to lower
resolution\\
Spectral calibration & Stability set by the velocity error budget & Required
for line-centroid and radial-velocity comparison across epochs\\
ToO response & Planning goal $\simeq2$ hr, fallback $<4$ hr, baseline
$<12$ hr & Records early kilonova and rapid accretion-state evolution\\
Photometric cadence & Sub-minute for selected targets & White-dwarf
pulsations and ultracompact-binary variability\\
Spectroscopic cadence & Minutes to days, event dependent & Separates orbital
modulation from outburst evolution\\
Roll-angle coverage & Two with a scene-dependent third orientation & Controls
overlapping traces without imposing unnecessary ToO overhead\\
PSF stability & Calibrated over each sequence & Enables line-flux extraction
and precise time-series subtraction\\
\bottomrule
\end{tabularx}
\end{table}

\subsection{Spectroscopic ETC implementation}

The \texttt{mission\_etc} package implements the initial spectroscopic ETC.
The count-rate and noise calculation resides in
\texttt{slitless\_etc.py}. The package provides separate modes for imaging,
isolated-line spectroscopy, template spectroscopy, and validation. The
compact-object calculation invokes only the spacecraft modes. A separate
cooled mid-infrared calculator supports comparisons with external ground
facilities and does not represent a 3.5-m spacecraft instrument. Separating
the modes prevents an AB magnitude, an integrated line flux, and a calibrated
$F_\lambda$ spectrum from entering the same calculation.

The template mode reads an observed-frame spectrum in physical flux units.
The calculation convolves the spectrum with a constant-$R$ Gaussian
line-spread function and samples the result on detector pixels. Source counts,
zodiacal and telescope backgrounds, dark current, read variance, and
contamination counts remain separate outputs. Each slitless detector pixel
receives the diffuse background integrated over the complete order-sorting
band. Relative flux calibration and contamination-model residuals are treated
as correlated terms when pixels are combined. The correlated terms do not
decrease as the square root of the number of binned pixels.

The current $R=5000$ settings cover 4300--5100\,\AA,
6200--6900\,\AA, and $1.0$--$1.5\micron$. Two $R=1000$ settings cover
$0.4$--$1.0\micron$ and $1.0$--$1.5\micron$ for broad transient spectra. The
implemented settings test the compact-object diagnostics and the
resolving-power trade. The current settings do not yet establish continuous
sensitivity over the adopted $0.2$--$1.5\micron$ envelope. Separate ultraviolet throughput and
detector models are required below $0.4\micron$. Additional order-sorting
settings are required between the present $R=5000$ bands.

The compact-red reference calculation yields a point-source
$5\sigma$ H$\alpha$ line limit of
$1.48\times10^{-17}\,\mathrm{erg\,s^{-1}\,cm^{-2}}$ in 1800\,s. The
6200--6900\,\AA\ band gives $3.24\times10^3$ sky electrons in a
2.55-pixel line footprint under the adopted background. The inverse
calculation recovers ${\rm S/N}=5$ to numerical precision. The calculation
uses proposal-level throughput and detector requirements rather than measured
flight performance. The line limit and background count will be revised after
measurements of component transmission, detector noise, and the point-source
line-spread function.

The ETC returns S/N per detector pixel and per combined spectral bin. The ETC
also returns line-flux uncertainty, centroid precision, a
wavelength-calibration floor, and the second velocity moment. A phase-smearing
diagnostic limits each exposure to a specified fraction of the binary period.
A broadband limiting magnitude alone is not an adequate compact-object
feasibility metric.

For an extraction aperture containing $N_{\rm pix}$ pixels, a reference
counting-noise expression is
\begin{equation}
 {\rm S/N} =
 \frac{N_{\rm src}}
 {\left[N_{\rm src}+N_{\rm pix}
 (N_{\rm sky}+N_{\rm dark}+n_{\rm read}\sigma_{\rm read}^2)
 +\sigma_{\rm cont}^2\right]^{1/2}},
 \label{eq:snr_budget}
\end{equation}
where $N_{\rm src}$ is the source count in the fitted spectral bin.
$N_{\rm sky}$ and $N_{\rm dark}$ are the sky and dark counts per pixel.
The number of reads is $n_{\rm read}$ and the read noise is
$\sigma_{\rm read}$. The contamination term $\sigma_{\rm cont}$ includes the
uncertainty of overlapping traces. In slitless data the sky term is set by the
full band-limiting filter recorded by each pixel. Increasing the native
dispersion may therefore reduce the S/N of a broad feature even when the final
spectrum is binned.

For a Gaussian line with integrated flux $F_{\rm line}$ and uncertainty
$\sigma_F$, the line detection statistic is
\begin{equation}
  {\rm S/N}_{\rm line} = \frac{F_{\rm line}}{\sigma_F},
  \qquad
  \sigma_{v,\rm stat} \propto
  \frac{c}{R\,{\rm S/N}_{\rm line}}.
  \label{eq:line_precision}
\end{equation}
The proportionality factor depends on the line profile and sampling. The
proposal should quote the full simulated value rather than assume that the
simple scaling is exact.

The total velocity error must also include a calibration floor,
\begin{equation}
 \sigma_v^2=\sigma_{v,\rm stat}^2+\sigma_{v,\rm wave}^2+
 \sigma_{v,\rm scene}^2+\sigma_{v,\rm model}^2 ,
 \label{eq:velocity_budget}
\end{equation}
where the additional terms arise from the wavelength solution, source
position and contamination, and the adopted line model. The requirement on
spectral stability should specify a maximum value for each error term. The
phrase
``stable wavelength solution'' is not a verifiable requirement by itself.

Mission-time estimates require the same quantitative treatment. The exact
integrations in Table~\ref{tab:target_selection} are planning values until
Equation~\ref{eq:snr_budget} is evaluated for the quiescent and outburst
magnitudes of every benchmark. The proposal should publish a matrix of
continuum S/N, line-flux S/N, and overlap survival probability for each source
class. That matrix determines which candidates remain in the baseline and
which become optional targets.

\subsubsection{Observed-template reference cases}

The observed-template calculation records the origin and native resolving
power of each spectrum. The line-spread calculation applies only the Gaussian
convolution required between the native resolution and the selected mission
setting. The calculation does not sharpen a template with lower resolving
power. Each template record retains the epoch, source state, wavelength frame,
flux uncertainty, citation, and public data URL.

\begin{figure}[H]
\centering
\includegraphics[width=0.99\textwidth]{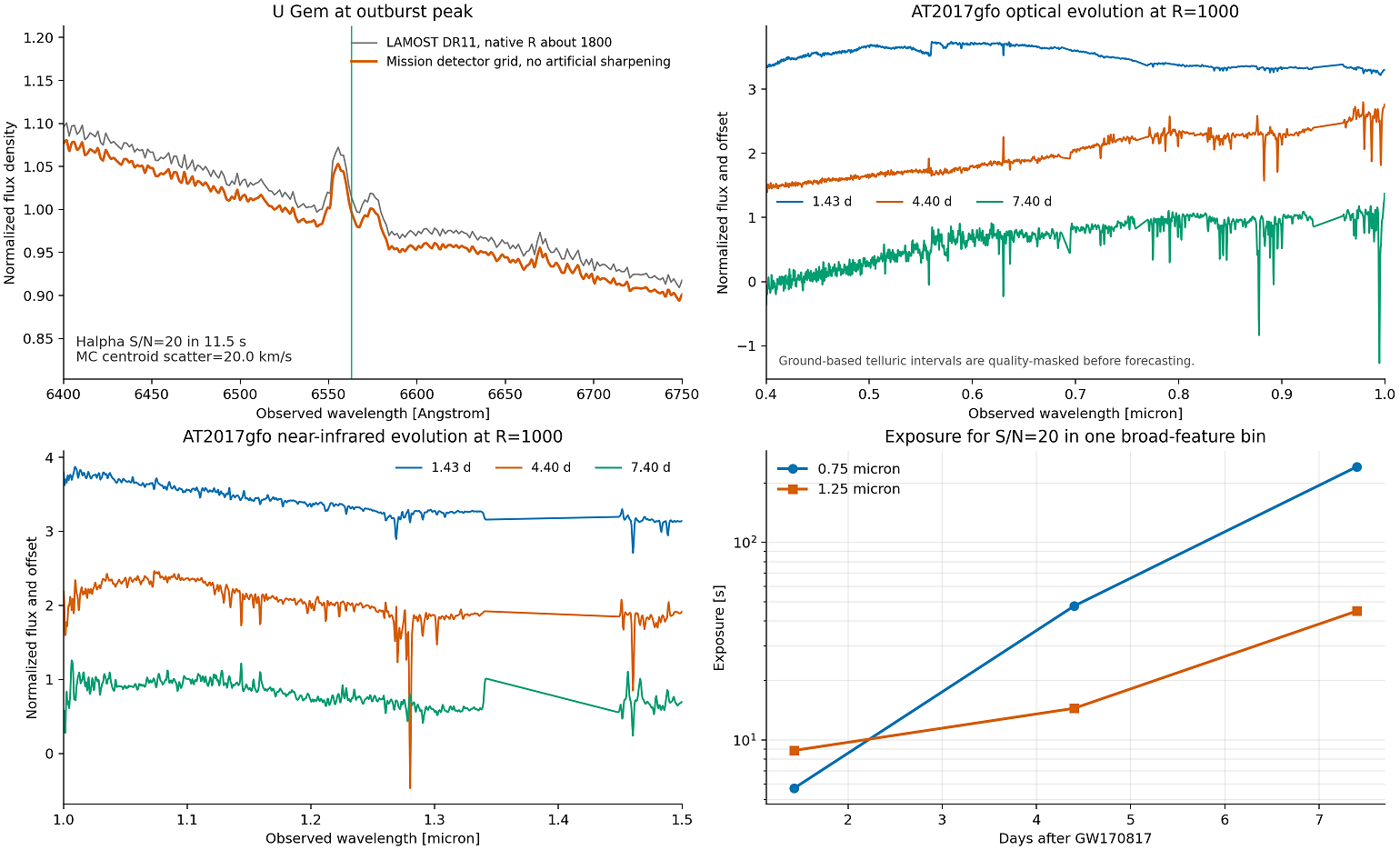}
\caption{Observed-template ETC reference cases. The upper left panel compares
the public LAMOST U~Gem spectrum with the detector-sampled calculation at the
native effective resolution. The other spectral panels use the public
ENGRAVE X-shooter sequence of AT~2017gfo. Known telluric intervals were
excluded before the broad-feature calculation because the ground-based
correction residuals are not intrinsic kilonova structure. Straight segments
across the excluded intervals represent interpolation. The lower right panel
shows the exposure required for an integrated S/N of 20 in a 150\,\AA\
optical interval and a 250\,\AA\ near-infrared interval. All exposure values
use proposal-level throughput and detector assumptions.}
\label{fig:compact_etc_reference}
\end{figure}

Figure~\ref{fig:compact_etc_reference} presents reference calculations based
on the LAMOST spectrum of U~Gem at the peak of its 2012 December outburst
\cite{zhi2020} and the ENGRAVE reduction of the AT~2017gfo X-shooter sequence
\cite{pian2017,smartt2017,engrave2019}. The U~Gem H$\alpha$ line reaches an
integrated S/N of 20 in 11.5\,s under the adopted instrument assumptions. The
native LAMOST resolving power is approximately 1800. The result constrains
counting statistics and broad line structure but does not validate
$R=5000$ double-peak recovery.

Three hundred detector-noise realizations of the U~Gem case gave a line
centroid bias of $1.6\,\kms$ and a scatter of $20.0\,\kms$. The median local
Fisher estimate was $6.2\,\kms$. The difference shows that a moment estimator
understates the uncertainty of the structured line profile. The next
validation stage must fit the complete profile and must report the empirical
covariance.

The AT~2017gfo calculation uses one broad-feature interval in each channel.
An integrated S/N of 20 at $0.75\micron$ requires 5.7, 47.7, and 241.5\,s at
1.43, 4.40, and 7.40 days after GW170817. The corresponding $1.25\micron$
exposures are 8.9, 14.5, and 45.0\,s. The exposure sequence demonstrates the
fading-rate calculation and the optical to near-infrared color evolution. The
values do not include alert latency, slew time, trace overlap, or a measured
flight throughput.

\subsubsection{Multi-orientation injection and recovery}

The multi-orientation calculation tests spectral confusion rather than
isolated-source counting statistics. The direct-image prior contains U~Gem and
two nearby field sources. Each source uses the public LAMOST U~Gem outburst spectrum with
an independent velocity and flux scale. The target receives an injected
velocity of $120\,\kms$. The forward calculation disperses the three sources
onto detector images and adds source shot noise, diffuse background, dark
current, and read noise.

The comparison fixes the total science exposure at 60\,s. The single
orientation receives the full 60\,s. The multi-orientation sequence receives
20\,s at each of $0^\circ$, $60^\circ$, and $120^\circ$. A common
direct-image position prior defines the source coordinates. The extraction
fits all source spectra simultaneously and then fits the recovered U~Gem
spectrum with the complete H$\alpha$ profile. The profile fit includes a
velocity shift, an amplitude, and a linear continuum.

\begin{figure}[H]
\centering
\includegraphics[width=0.99\textwidth]{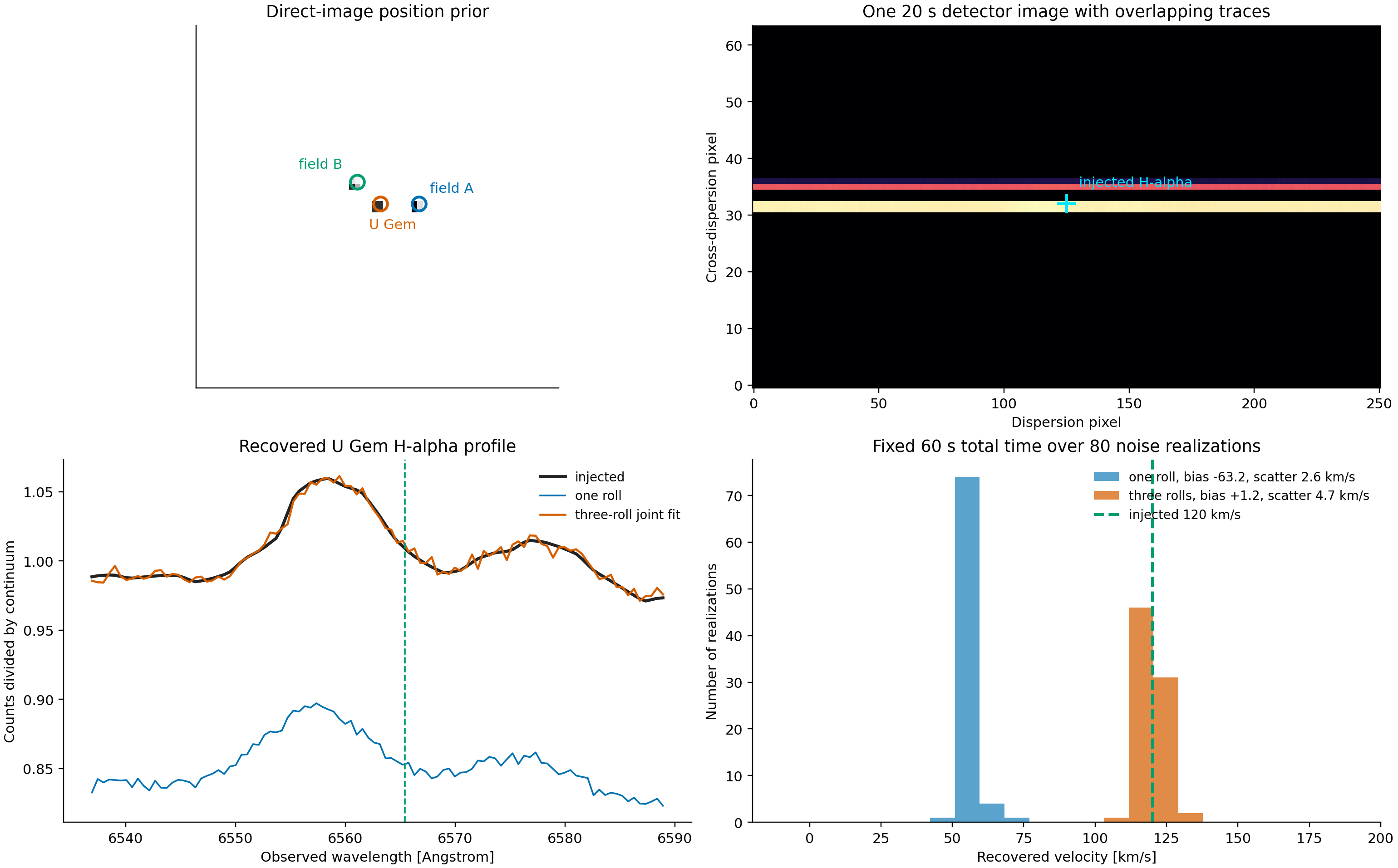}
\caption{Fixed-time slitless injection and recovery using the public LAMOST
U~Gem spectrum. The upper panels show the direct-image position prior and one
20\,s detector exposure. The lower left panel compares the injected
H$\alpha$ profile with the single-orientation and joint three-orientation
extractions. The lower right panel shows the recovered velocity for 80 noise
realizations. Both observing strategies use 60\,s of total science exposure.
The LAMOST spectrum has an effective resolving power near 1800. The
experiment validates multi-orientation trace separation and does not validate
double-peak recovery at $R=5000$.}
\label{fig:compact_scene_recovery}
\end{figure}

Figure~\ref{fig:compact_scene_recovery} shows the result for 80 independent
detector-noise realizations. All 80 fits converged in both observing
strategies. The single-orientation fit gives a median velocity of
$56.6\,\kms$. The bias is $-63.2\,\kms$ and the empirical scatter is only
$2.6\,\kms$. The small scatter does not demonstrate an accurate measurement
because the contaminating trace lies along the target trace. Additional
counts do not remove the geometric degeneracy.

The three-orientation fit gives a median velocity of $119.4\,\kms$. The bias
falls to $+1.2\,\kms$ and the empirical scatter is $4.7\,\kms$. Dividing the
exposure among three orientations modestly increases the random error while
the different projected separation removes the much larger systematic error.
The median diagonal-Fisher estimate is $2.8\,\kms$ for the joint fit. The
empirical scatter must set the velocity requirement because the diagonal
estimate omits covariance among overlapping spectra.

\section{Network Complementarity and Limitations}

The compact-object program requires external facilities for discovery,
high-energy measurements, and joint physical inference.

\begin{table}[H]
\centering
\caption{Division of labor among the external facilities retained in the
compact-object baseline. Mid-infrared measurements are assigned to targeted
ground-based follow-up and are not a 3.5ST observing mode.}
\label{tab:network}
\begin{tabularx}{\textwidth}{@{}P{34mm} Y Y@{}}
\toprule
\textbf{Facility class} & \textbf{Primary information} & \textbf{Role of the
3.5 m space telescope}\\
\midrule
Gravitational-wave detectors & Binary masses, distance, and tidal information &
Rapid electromagnetic localization and spectral follow-up\\
X-ray and gamma-ray missions & Jet, disk, and high-energy trigger &
Optical/NIR response and host identification\\
Wide-field optical surveys & Discovery and large-area cadence &
High-quality repeated spectra and rapid response\\
Large ground telescopes & Very high-resolution spectroscopy &
Space-stable moderate-resolution time series and early response\\
\bottomrule
\end{tabularx}
\end{table}

A wide gravitational-wave localization region makes spectroscopy inefficient
until an optical counterpart is identified. The mission therefore responds to
a localized counterpart rather than tiling the full gravitational-wave
probability map with $R=5000$ spectra. Optical and near-infrared data alone
cannot separate kilonova ejecta mass, composition, opacity, geometry, and
viewing angle. The inference therefore requires an external
gravitational-wave posterior and a radiative-transfer model.

The telescope also does not replace X-ray, gamma-ray, or high-resolution
spectroscopic facilities. The moderate-resolution time series measures a
stable lower-energy response that is difficult to obtain from the ground at
the same cadence. High-resolution spectra remain necessary for weak metal
lines and detailed abundance work. Extended hosts and unrelated sources may
overlap a slitless trace. Every released spectrum must therefore include a
scene-quality flag and the posterior contamination fraction.

White-dwarf interior parameters provide a final example of the division of
labor. The spectrum determines the atmospheric boundary condition, and the
imaging measures the pulsation frequencies. Stellar models infer the interior
from both measurements. The science requirements and released data products
must distinguish the measured atmospheric and pulsation quantities from the
model-dependent interior parameters.

\section{Five-Year Reference Allocation}
\label{sec:compact_schedule}

The shared mission scheduler reads the target classes, exposure times,
cadences, solar-elongation limits, maintenance periods, and Director weeks
from the observing database. The scheduler then assigns every program to a
common 260-week capacity model. The current reference solution reserves 618.3 hr for
Volume V. The allocation equals 1.41 per cent of the five-year wall clock.
Table~\ref{tab:compact_five_year} gives the program totals.

\begin{table}[H]
\centering
\caption{Database-driven five-year reference allocation for Volume V. The
time values include the programmed science integrations and the reserved ToO
capacity represented in the shared scheduler.}
\label{tab:compact_five_year}
\begin{tabularx}{\textwidth}{@{}P{42mm} C{22mm} C{24mm} Y@{}}
\toprule
\textbf{Program} & \textbf{Mission years} & \textbf{Time [hr]} &
\textbf{Scheduling purpose}\\
\midrule
Dwarf-nova intensive sample & 1--2 & 31.5 & Dense sequences for bright
systems during selected outbursts\\
Dwarf-nova population sample & 1--2 & 60.0 & Repeated state sampling across
the controlled catalog sample\\
AM CVn monitoring & 2--3 & 90.0 & State monitoring and one
phase-resolved sequence for selected ultracompact binaries\\
Compact-object ToO reserve & 1--5 & 436.8 & One per cent of the mission wall
clock retained for merger counterparts and accretion transients\\
\midrule
\textbf{Volume V total} & \textbf{1--5} & \textbf{618.3} &
\textbf{1.41 per cent of the five-year wall clock}\\
\bottomrule
\end{tabularx}
\end{table}

\begin{figure}[H]
\centering
\includegraphics[width=0.99\textwidth]{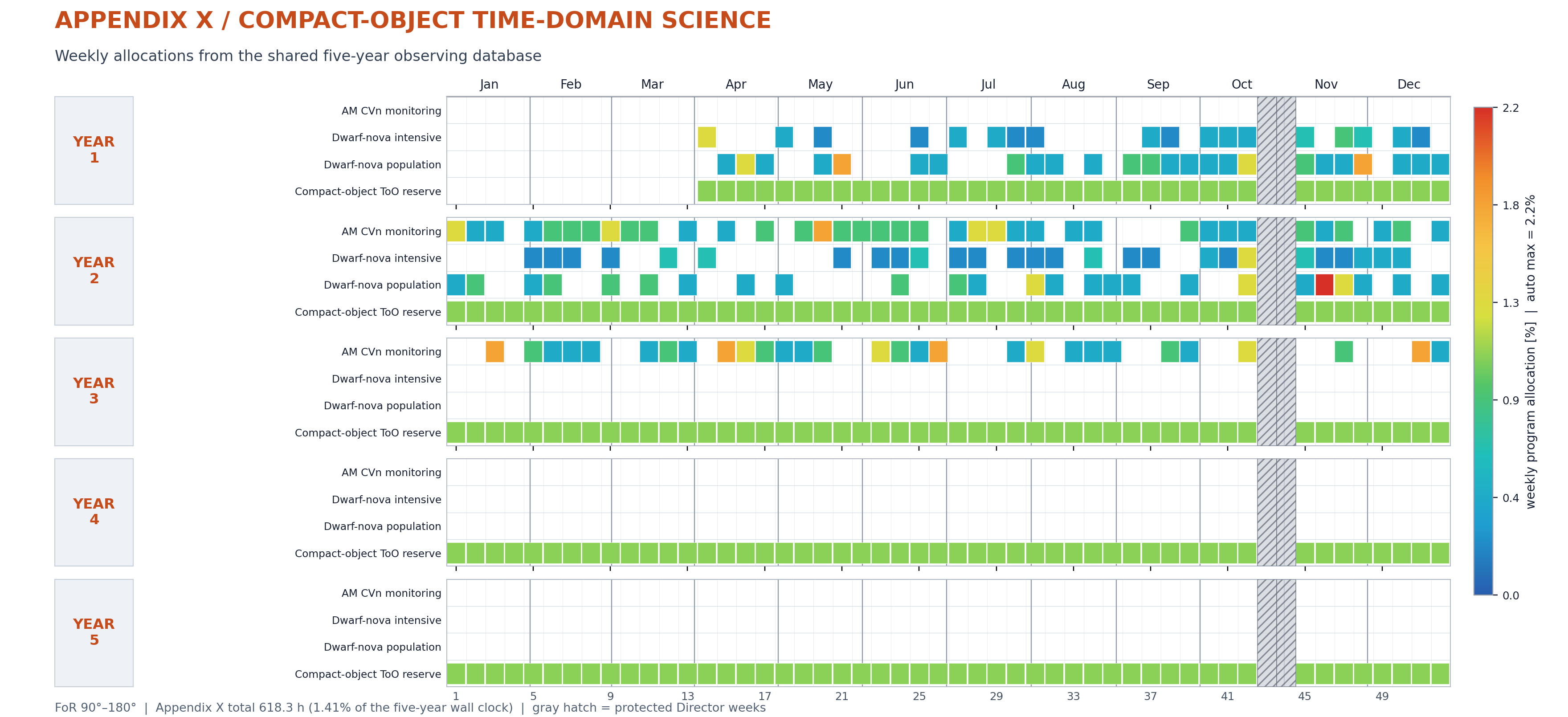}
\caption{Weekly Volume V allocation generated from the shared observing
database. Color gives the fraction of one 168-hr week assigned to each
program. The color scale uses the largest Volume V allocation and the gray
hatching marks the protected Director weeks. The ToO row represents reserved
capacity rather than preselected targets.}
\label{fig:compact_five_year}
\end{figure}

Figure~\ref{fig:compact_five_year} shows the weekly assignments. Routine
science begins in Year 1 week 14 after commissioning and performance
acceptance. The scheduler excludes Director weeks 43 and 44 in every year and
tests fixed targets against a solar elongation of $90^\circ$--$180^\circ$.
The compact-object ToO time remains distributed over otherwise eligible weeks
until an alert supplies a coordinate, a discovery time, and a predicted
brightness. The figure therefore demonstrates available mission capacity. The
allocation does not predict the number of gravitational-wave counterparts or
accretion transients that will satisfy the trigger criteria.

\section{Required Validation Before Proposal Baseline}

The compact-object program can support a mission-level performance claim only
after the validation products below meet quantitative acceptance tests.

The first validation product is an ETC grid spanning magnitude, line flux,
source state, background, detector channel, and roll-dependent contamination
for kilonovae, dwarf novae, white dwarfs, and accreting neutron-star or
black-hole binaries. The grid must be based on observed spectral templates.
The second product is a slitless forward simulation using direct-image
priors, detector sampling, multiple dispersion orientations, time-dependent
source flux, and unrelated field sources. A blinded recovery test should then
measure bias and covariance for injected line profiles and known velocities.

Operational feasibility requires a separate scheduler simulation. The
simulation must begin with a realistic alert stream and must include
classification latency, field of regard, solar avoidance, slew time, roll
availability, guide-star acquisition, downlink, and interrupted visits. The
simulation must return a five-year yield that distinguishes public triggers,
identified counterparts, observed targets, spectra with usable line S/N, and
science-grade multi-epoch sequences.

Two class-specific demonstrations complete the validation. Archival spectra
and light curves of dwarf novae with known orbital periods should test the
recovery of peak separation, wings, velocity modulation, and transitions
between states
after convolution to the proposed instrument. Injected white-dwarf light
curves should quantify the frequency bias produced by cadence, gaps, and
photometric noise. The class-specific demonstrations connect each proposed
measurement to a traceable performance requirement.

\begin{xresultbox}{orange}
\textbf{Recommended proposal position.}
The compact-object program is scientifically credible as a point-source
optical and near-infrared time-domain program. Mission feasibility remains
conditional on the ETC, scene-recovery, scheduler, and yield demonstrations
defined above. The $R\simeq5000$ bands are most valuable for accretion-disk
profiles and should provide lower-resolution binned products for broad
transients. The science claims require joint inference with gravitational-wave,
high-energy, and radiative-transfer information. The program does not require
a mid-infrared instrument and does not claim a direct equation-of-state or
interior measurement from one spectrum.
\end{xresultbox}

\section{Summary}

The adopted compact-object baseline uses a $0.2$--$1.5\micron$ space-based
slitless spectrograph. The $R\simeq1000$ setting measures broad transient
spectra, while selectable $R\simeq5000$ bands measure precision line profiles.
The common instrument study extends calibrated throughput to $2.70\micron$
and retains $3.0\micron$ as an operational band-edge goal. The small angular
size of an isolated compact object preserves the native spectral resolution.
Direct images, multiple orientations, and a scene model remain necessary for
separating overlapping traces. After the ETC verifies the required S/N, the
spectra measure velocities, ionization, outflows, disk kinematics, transitions
between states, and transient spectral evolution.

Rapid optical and near-infrared follow-up of mergers containing at least one
neutron star provides the strongest transient case. Dwarf novae provide a
repeatable population for testing accretion-disk instability and binary
dynamics with line-profile sequences. For white dwarfs, atmospheric
spectroscopy and asteroseismology jointly probe cooling, crystallization, and
remnant planetary systems over a long baseline. The three experiments use the
same detector but require different cadence, resolving power, calibration, and
mission operations.

\appendixXfinish

\let\APPENDIXXBOOK\undefined

\clearpage
\definecolor{Blue1}{HTML}{1FABD5}
\definecolor{Blue2}{HTML}{1D8DB0}
\definecolor{Blue3}{HTML}{116E8A}
\backmatter
\MakeBackCover
\end{document}